\documentclass{CSML}
\pdfoutput=1

\usepackage{lastpage}
\newcommand{\dOi}{14(2:3)2018}
\lmcsheading{}{1--\pageref{LastPage}}{}{}%
{Sep.~14, 2016}{Apr.~25, 2018}{}
\usepackage{hyperref}
\hypersetup{hidelinks}
\usepackage{tikz}
\usepackage[english]{babel}
\usepackage[utf8]{inputenc}
\usepackage{amsfonts,amsmath,amssymb,amsthm,latexsym}

\usetikzlibrary{arrows}
\usetikzlibrary{petri}
\usetikzlibrary{backgrounds}
\usetikzlibrary{decorations.pathmorphing}
\usetikzlibrary{calc}
\tikzstyle{vertex}=[circle,draw=blue!50,fill=blue!20,thick]
\tikzstyle{sipka}=[->,shorten <=1pt,>=angle 90,semithick]
\tikzstyle{forced}=[->,shorten <=1pt,>=angle 90,semithick,dashed]
\tikzstyle{bigbad}=[line width=3pt]
\tikzstyle{every token}=[draw=blue!50,fill=blue!20,thick]

\let\phi\varphi
\def\zet{{\mathbb Z}}

\def\Im{\operatorname{Im}}

\newcommand\comma{,\,}

\DeclareMathOperator\Conn{Conn}

\let\subset\subseteq
\def\compNP{{\textsf{NP}}}
\def\compNL{{\textsf{NL}}}
\def\compSL{{\textsf{SL}}}
\def\compL{{\textsf{L}}}
\newcommand\compModL[1]{\textsf{Mod}_#1\textsf{L}}
\def\compP{{\textsf{P}}}

\def\algA{{\mathbf{A}}}

\def\A{{\mathbb{A}}}
\def\B{{\mathbb{B}}}

\newcommand\C{{\mathcal C}}
\newcommand\D{{\mathcal D}}
\newcommand\Arrr{{\mathcal R}}

\DeclareMathOperator\CSP{CSP}

\DeclareMathOperator\symprogram{{\mathcal P}}
\DeclareMathOperator\proves{\vdash}

\def\CSP{\operatorname{CSP}}

\def\A{{\mathbb A}}
\def\B{{\mathbb B}}
\def\en{{\mathbb N}}

\def\CSP{\operatorname{CSP}}

\let\epsilon\varepsilon

\theoremstyle{plain}
\newtheorem*{thm*}{Theorem}

\title{$n$-permutability and linear Datalog implies symmetric Datalog}
\author[A.~Kazda]{Alexandr Kazda}
\address{Department of Algebra Charles University Prague, Czech Republic}
\email{alex.kazda@gmail.com}
\begin{document}

\begin{abstract}
  \noindent We show that if $\mathbb A$ is a core relational structure such
  that CSP($\mathbb A$)
  can be solved by a linear Datalog program, and $\mathbb A$ is $n$-permutable
  for some $n$,
  then CSP($\mathbb A$) can be solved by a symmetric Datalog program (and thus
  CSP($\mathbb A$)
  lies in deterministic logspace). At the moment, it is not known for which
  structures $\mathbb A$ will CSP($\mathbb A$) be solvable by a linear Datalog program. However,
  once somebody obtains a characterization of linear Datalog, our result
  immediately gives a characterization of symmetric Datalog.
\end{abstract}
\maketitle
\section{Introduction}%{{{1

In the last decade, algebraic methods have led to much progress in classifying
the complexity of the non-uniform Constraint Satisfaction Problem (CSP). The
programming language Datalog, whose origins lie in logic programming and
database theory, has been playing an important role in describing the complexity
of CSP since at least the classic paper of T. Feder and M.
Vardi~\cite{feder-vardi}, where Feder and Vardi used Datalog to define CSPs of
bounded width. In an effort to describe the finer hierarchy of CSP complexity,
V. Dalmau~\cite{dalmau-linear-datalog-pw-duality} asked which CSPs can be
solved using the weaker language of linear Datalog, and later L. Egri, B.
Larose and P. Tesson~\cite{egri-larose-tesson-symmetric-datalog} introduced
the even weaker symmetric Datalog. 

We want to show that if $\CSP(\A)$ can be solved by a linear Datalog program
(alternatively, has bounded pathwidth duality) and $\algA$ generates an
$n$-permutable variety for
some $n$, then $\CSP(\A)$ can be solved by a \emph{symmetric} Datalog program
(and so lies in $\compL$). While this yields an ``if and only if'' description
of symmetric Datalog, it is not a perfect characterization -- describing the structures $\A$
such that $\CSP(\A)$ is solvable by linear Datalog is an open problem. However, once CSPs for which
linear Datalog works are classified, we will immediately get an equally good
classification of symmetric Datalog CSPs.

In particular, should it turn out that admitting only the lattice and/or
Boolean tame congruence types implies bounded pathwidth duality, we
would have a neat characterization of problems solvable by symmetric Datalog:
It would be the class of problems whose algebras omit all tame congruence theory
types except for the Boolean type (we go into greater detail about tame
congruence theory in Preliminaries and Conclusions).

Our result is similar to, but incomparable with what V. Dalmau and B. Larose
have shown~\cite{dalmau-larose-symmetric-datalog}: Their proof shows that
2-permutability plus being solvable by Datalog implies solvability by symmetric
Datalog. We require both less ($n$-permutability for some $n$ as opposed to
2-permutability) and more (linear
Datalog solves $\CSP(\A)$ as opposed to Datalog solves $\CSP(\A)$).

Our proof strategy is this: First we show in Section~\ref{secStacking} how we
can use symmetric Datalog to derive new instances from the
given instance. Basically, we show that we can run a smaller symmetric Datalog
program from inside another. This will later help us to reduce ``bad'' CSP instances to
a form that is easy to deal with. Then, in Section~\ref{secNPermutable} we
introduce path CSP instances and show how $n$-permutability restricts the kind
of path instances we can encounter. We use this knowledge in
Section~\ref{secSymDLPaths} to show that for any variety $n$-permutable $\A$, there is a
symmetric Datalog program that decides path instances of $\CSP(\A)$. Finally, in
Section~\ref{secFinale} we use linear Datalog to go from solving path instances
to solving general CSP instances and finish our proof.

When writing this paper, we were mainly interested in ease of exposition, not
in obtaining the fastest possible algorithm. We should therefore warn any
readers hoping to implement our method in practice that the size of our
symmetric Datalog program grows quite quickly with the
size of $\A$ and the number of Hagemann-Mitschke terms involved. The main
culprit is Lemma~\ref{lemBraidExists} that depends on Ramsey theory.
%}}}
\section{Preliminaries}%{{{1 \label{secPreliminaries} 
All numbers in this paper
are integers (most of them positive). If $n$ is a positive integer and $a,b$ are
integers, we will use the notation
$[n]=\{1,2,\dots,n\}$ and the notation
$[a,b]=\{i\in \zet\colon a\leq i\leq b\}$ (and variants such as $[a,b)=\{i\in
\zet\colon a\leq i< b\}$).

We will be talking quite a bit about tuples -- either tuples of elements of $A$ or
tuples of variables. We will treat both cases similarly: An $n$-tuple on $Y$
is a mapping $\sigma\colon [n]\to Y$. We will denote the length
of the tuple $\sigma$ by $|\sigma|$, while $\Im \sigma$ will be the set of elements used in
$\sigma$. Note that if e.g. $\sigma=(x,x,y)$, we can have $|\sigma|>|\Im
\sigma|$.

A \emph{relation} on $A$ is any $R\subset A^X$ where $X$ is some (finite)
set. The \emph{arity} of $R$ is the cardinality of $X$. Most of the time, 
we will use $X=[n]$ for some $n\in\en$ and write simply $R\subset A^n$.

When $R\subseteq A^n$ is an $n$-ary relation and $\sigma=(a_1,\dots,a_n)$ is an
$n$-tuple, we will often write $R(\sigma)$ instead of $(a_1,\dots,a_n)\in R$.
Given a mapping $f:A\to B$ and an $n$-tuple $\sigma\in A^n$,
we will denote by $f(\sigma)$  the $n$-tuple
$(f(\sigma(1)),\dots,f(\sigma(n)))\in B^n$.

\subsection{Algebras and relational structures}
We will be touching some concepts from universal algebra that
would deserve a more detailed treatment than what we will provide
here. See~\cite{uabook} for an introduction to universal algebra.

A \emph{relational structure} $\A$ consists of a set $A$ together with a family
$\Arrr$ of relations on $A$, which we call \emph{basic relations} of $\A$.  In
this paper, we will only consider finite relational structures with finitely
many basic relations. We will not allow nullary relations or relations of
infinite arity.

An \emph{$n$-ary operation} on $A$ is any mapping $t\colon A^n\to A$.  We say
that an $n$-ary operation $t$ \emph{preserves} the relation $R$ if for all
$r_1,\dots,r_n\in R$ we have $t(r_1,r_2,\dots,r_n)\in R$ (where
$t(r_1,\dots,r_n)$ is the tuple we obtain by applying $t$
componentwise to $r_1,\dots, r_n$). Given a relational structure $\A$, an
$n$-ary operation $t$ on $A$ is a \emph{polymorphism} of $\A$ if $t$ preserves
all basic relations of $\A$.

An \emph{algebra} $\algA$ consists of a base set $A$ on which acts a set of
\emph{basic operations} of $\algA$. We can compose basic operations of $\algA$
to get more operations. A \emph{term} in $\algA$ is a correctly formed string
that consists of variables and basic operation symbols of $\algA$ (as well
as parentheses and commas) and describes a meaningful composition of
operations. For example, if $\algA$ has the ternary basic operation $r$, then
``$r(x_3,r(x_1,x_1,x_2),x_4)$'' is a term in $\algA$ that describes the 4-ary operation
(with variables $x_1,\dots,x_4$) we
can get by composing $r$ with itself in a particular way. An algebra is
\emph{idempotent} if for any term operation $t$ in $\algA$ and any $a\in A$ we have
$t(a,\dots,a)=a$. 

The \emph{algebra of polymorphisms of $\A$} is the algebra with the universe $A$
whose set of operations consists of all polymorphisms of $\A$. We will use
the shorthand $\algA$ for this algebra.

A \emph{congruence} $\alpha$ on an algebra $\algA$ is any binary
equivalence relation that is preserved by all operations of
$\algA$.

The \emph{relational clone} of $\A$ is the set of all relations
on $A$ that can be defined from the basic relations of $\A$ by
primitive positive definitions -- formulas that only use
conjunction, existential quantification and symbols for
variables. We will sometimes call members of the relational clone
of $\A$ \emph{admissible relations of $\A$}. The importance of
the relational clone comes from the fact that $\algA$ preserves
precisely all relations on $A$ that belong in the relational
clone of $\A$~\cite{bodnarchuk, geiger}.

A \emph{variety} is a class of algebras sharing the same
signature (the same basic operation symbols and arities of basic
operations) that is closed under taking subalgebras, products and
homomorphic images. If $\algA$ is an algebra, then the variety
generated by $\algA$ is the smallest variety that contains
$\algA$, or equivalently the class of all homomorphic images of
subalgebras of powers of $\algA$.

Since all algebras in a given variety have the same signature, it
makes sense to talk about term operations of a variety. We will be using one
particular set of such operations, called Hagemann-Mitschke terms, in our proofs.

Let us fix a positive integer $n$. We say that a variety $V$ is 
(congruence) $n$-permutable if for any algebra $\algA$ in $V$
and any pair of congruences $\alpha$, $\beta$ of $\algA$ it is true that
\[
  \alpha\vee\beta=\alpha\circ\beta\circ\alpha\circ\dots
\]
with $n-1$ composition symbols on the right side (in particular, 2-permutable
is the same thing as congruence permutable).

A standard free algebra argument gives us that $V$ is $n$-permutable if and
only if we can find idempotent terms $p_0,p_1,\dots,p_n$ in $V$ such that
\begin{align*}
  x&=p_0(x,y,z),\\
  p_i(x,x,y)&=p_{i+1}(x,y,y)\quad \text{for all $i=1,2,\dots,n-1$},\\
  p_n(x,y,z)&=z.
\end{align*}

The above terms are called Hagemann-Mitschke terms and were first obtained
in~\cite{hagemann-mitschke}.

If the algebra of polymorphisms of a relational structure $\A$ generates an
$n$-permutable variety, i.e. if there are Hagemann-Mitschke operations
$p_0,p_1,\dots,p_n$ in $\algA$, we say simply that $\A$ is
variety $n$-permutable (the ``variety'' prefix is here to emphasize that the
whole variety of $\algA$, not just $\algA$ itself, needs to have $n$-permutable congruences). There are several other
conditions that connect the behavior of congruences in a variety with the
variety having certain term operations. We mention (without going into details) congruence
distributivity and congruence semidistributivity in the next section.

\subsection{Constraint Satisfaction Problem}
Let us fix a relational structure $\A=(A,\Arrr)$ and define the 
non-uniform \emph{Constraint Satisfaction Problem} with the right side $\A$, or $\CSP(\A)$
for short. This problem can be stated in several mostly equivalent ways 
(in particular, many people prefer to think of $\CSP(\A)$ as a question about
homomorphisms between relational structures). We define $\CSP(\A)$ in the
language of logical formulas.
\begin{defi}\label{defCSP}
An instance $I=(V,\C)$ of $\CSP(\A)$ consists of a set of variables $V$ and a
set of constraints $\C$. Each constraint is a pair $(\sigma, R)$ 
where $\sigma\in V^n$ is the \emph{scope} of the
constraint and $R\in \Arrr$ is the \emph{constraint relation}. A
\emph{solution} of $I$ is a mapping $f\colon V\to A$ such that for all
constraints $(\sigma, R)\in\C$ we have $f(\sigma)\in R$.

If $I$ is an instance, we will say that $I$ is \emph{satisfiable} if there
exists a
solution of $I$ and \emph{unsatisfiable} otherwise. The Constraint Satisfaction Problem 
with target structure $\A$ has as its input an instance $I$
of $\CSP(\A)$ (encoded in a straightforward way as a list of
constraints), and the output is the answer to the question ``Is $I$
satisfiable?''
\end{defi}

If $I=(V,\C)$ is an instance of $\CSP(\A)$, then any $\CSP(\A)$ instance
$J=(U,\D)$ with $U\subset V$ and $\D\subset \C$ is called a \emph{subinstance}
of $I$. It easy to see that if $I$ has an unsatisfiable subinstance then $I$
itself is unsatisfiable. If $U\subset V$, the subinstance of $I=(V,\mathcal C)$
induced by $U$ is the instance $I_{\upharpoonright U}=(U,\mathcal D)$  where $(\sigma,R)\in \mathcal D$ if and
only if $\Im \sigma\subset U$.

%A typical case of such ambiguity will be
%when $R$ on $A$ is an $n$-ary relation of $\A$, while $R(\sigma)$ for $\sigma\in V^n$ is the
%statement ``whenever $f\colon V\to A$ is a solution of a given CSP instance, then
%$f(\sigma)\in R$''. In this second sense, we will often call $R$ a predicate.
%In this formalism, an instance of $(V,\C)$ of $\CSP(\A)$ can be stated as
%``Decide if there exists an $f\colon V\to A$ such that $\wedge_{(\sigma,R)\in\C}
%R(\sigma)$ is true.''

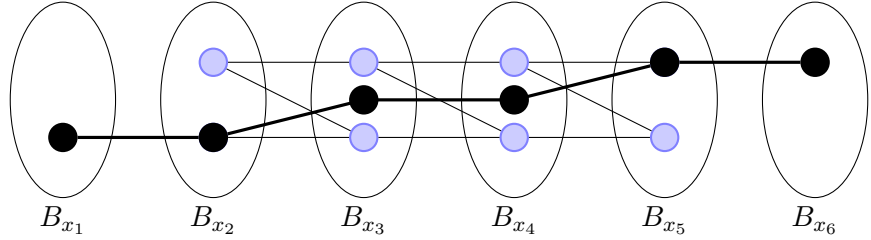
\begin{figure}
  \begin{center}
    \begin{tikzpicture}[bend angle=45,draw,auto, node distance=0.5cm]
      \node[vertex, black] at (0,-1cm) (a0){};
	\draw (0,-0.5cm) ellipse (0.7cm and 1.3 cm);

      \node[vertex, black] at (10cm,0cm) (a5){};
	\draw (10cm,-0.5cm) ellipse (0.7cm and 1.3 cm);
	\foreach \i in {1,2,3,4,5,6} {
	  \node at ($(2*\i cm,-2.1)+(-2,0)$) {$B_{x_\i}$};
	}
      \foreach \x/\i in {1/2,2/3,3/4,4/5} {
	\node[vertex] at (2*\x cm,0) (a\x){};
	\node[below of=a\x] (b\x){};
	\node[vertex, below of=b\x] (c\x){};
	\draw (b\x.center) ellipse (0.7cm and 1.3 cm);
      }
      \node[vertex, black] at (c1) {};
      \node[vertex, black] at (a4) {}; 
      \node[vertex,black] at (b2){};
      \node[vertex,black] at (b3){};
      \foreach \x/\y in
      {c1/c2,c2/c3,a1/c2,a2/c3,a3/c4,a1/a2,a2/a3,c3/c4,a3/a4}
         \draw (\x) edge (\y);
      \foreach \x/\y in {a0/c1,c1/b2,b2/b3,b3/a4,a4/a5}
      \draw[very thick] (\x) edge (\y);
        \end{tikzpicture}
  \end{center}
  \caption{An example of microstructure with six variables $x_1,x_2,\dots,x_6$
    and five binary relations (instance solution in bold).}
  \label{figPotato}
\end{figure}

We can draw CSP instances whose constraints' arities are at most two as
\emph{microstructures} (also known as potato diagrams among universal
algebraists): For each variable $x$ we draw the set $B_x\subset A$ equal to the
intersection of all unary constraints on $x$. For each binary constraint we draw
lines joining the pairs of elements in corresponding sets.  A solution of the
instance corresponds to the selection of one element $b_x$ in each set $B_x$ in
such a way that whenever $\C=((x,y),R)$ is a constraint, we have $(b_x,b_y)\in
R$ (see Figure~\ref{figPotato} for an example). 

Obviously, $\CSP(\A)$ is always in the class
\compNP, since we can check in polynomial time whether a mapping $f\colon V\to A$ is
a solution. Had we let the structure $\A$ be a part of the input, the
constraint satisfaction problem would be \compNP-complete (it is easy to
encode, say, 3-colorability of a graph as a CSP instance). However, when one
fixes the structure $\A$, $\CSP(\A)$ can become easier.

A relational structure $\A$ is a~\emph{core} if any unary polymorphism $f\colon
\A\to\A$ is an automorphism (i.e. we cannot retract $\A$ to a smaller
relational structure). To classify the complexity of $\CSP(\A)$ for $\A$
finite, it is enough to classify cores, see~\cite[p. 142]{hell-nesetril}.

The value of the algebraic approach to $\CSP$ is that it connects the
complexity of $\CSP(\A)$ to the algebra of polymorphisms $\algA$. Let us
highlight one such important connection here: The tame congruence theory is a
tool that arose from the study of finite algebras in the 1980s. In a 
nutshell (see~\cite{hobby-mckenzie} for more),
the theory aims to connect properties of congruences in a variety
with the existence of certain term operations in that variety and with
the set of so-called \emph{types} that the variety admits (where ``admits''
means that a certain technical construction can, when run on a suitable algebra
from the variety, produce a given type). It turns
out~\cite{larose-tesson-hardness-csp} that we can get lower bounds on the
complexity of $\CSP(\A)$ from the set of types that the variety
generated by $\algA$ admits. Let us rephrase four out of the five parts of 
Theorem 4.1 from~\cite{larose-tesson-hardness-csp} to
include concrete problems that each type brings about:
\begin{thm}[B. Larose, P. Tesson; part 1 is due to A. Bulatov, P. Jeavons,
  and A. Krokhin~\cite{bulatov-krokhin-jeavons-CSP-and-algebra}]\label{thmTypeLB}
  Let $\A$ be a core relational structure (with finitely many basic relations) and let $V$ be the variety generated by
$\algA$. Then:
\begin{enumerate}
  \item If $V$ admits the unary type (type 1), then 3-SAT reduces to
    $\CSP(\A)$ and hence $\CSP(\A)$ is \compNP-hard,
  \item if $V$ admits the semilattice type (type 5), then HORN-3-SAT reduces
    to $\CSP(\A)$ and hence $\CSP(\A)$ is \compP-hard,
  \item if $V$ admits the affine type (type 2), then there is a prime $p$ such
    that the complement of the $p$-GAP problem
    (see~\cite{larose-tesson-hardness-csp} for definition) reduces to $\CSP(\A)$ and
    hence $\CSP(\A)$ is $\compModL{p}$-hard for some prime $p$,
  \item if $V$ admits the lattice type (type 4), then the directed
    unreachability problem (the complement of directed reachability)
    reduces to $\CSP(\A)$ and hence $\CSP(\A)$ is
    $\compNL$-hard.
\end{enumerate}
  All the reductions above are first order reductions
  (see~\cite{descriptive-complexity} for definition) and are very natural
  from the point of view of universal algebra.
\end{thm}
Notice that one type is missing from the above theorem: The Boolean type (type
3) only gives us very weak lower bounds. When $\algA$
generates a variety that admits the Boolean type only, the problem $\CSP(\A)$ 
is believed to be in \compL{} (see Conjecture~\ref{conjType3}).

\subsection{Datalog}%{{{2
The Datalog language offers a way to check the local consistency of CSP
instances. A Datalog program $P$ for solving $\CSP(\A)$ consists of a list of rules
of the form
\[
R(\rho)\leftarrow  S_1({\sigma_1})\comma
S_2({\sigma_2})\comma\dots \comma S_\ell({\sigma_\ell}),
\]
where $R, S_1,\dots,S_l$ are predicates and $\rho,{\sigma_1},\dots,
{\sigma_\ell}$ are sequences of variables (we will denote the set of all variables
used in the program by $X$). Some predicates of $P$ are designated as \emph{goal
predicates} (more on those later).

In general, the predicates can be symbols without any meaning, but in the
programs we are about to construct each predicate will correspond to a relation
on $A$, i.e. a predicate $S(x_1,x_2)$ would correspond to some $S\subset A^2$.
This will often get us in a situation where, say, the symbol $R$ stands at the
same time for a relation on $A$, a predicate of a Datalog program, and a
relation on the set $V$ of variables (see below). For the most
part, we will depend on context to tell these meanings of $R$ apart, but if
there is a risk of confusion we will employ the notation $R^A$ for
$R^A\subset A^n$, $R^P$ for predicates of $P$, and $R^V$ for $R^V\subset V^n$.

Given a Datalog program $P$ that contains predicates for all basic relations of
a relational structure $\A$, we can run $P$ on an instance $I=(V,\C)$ of
$\CSP(\A)$ as follows: For each $n$-ary predicate $R^P$ of $P$, we keep in
memory an $n$-ary relation $R^V\subset V^n$. Initially, all such relations are
empty. To load $I$ into the program, we go through $\C$ and for every
$(\sigma,R^A)\in\C$, we add $\sigma$ to $R^V$ (when designing $P$, we will always make sure that
there is a predicate $R^P$ for each basic relation $R^A$ of $\A$).

After this initialization, $P$ keeps adding tuples of $V$ into relations $R^V$
as per the rules of $P$: If we can assign values to variables so that the right
hand side of some rule holds, then we put the corresponding tuple into the left
hand side relation $R$. 

More formally, we say that $P(I)$ derives $R^V(\rho)$ for $\rho\in V^n$,
writing $P(I)\proves R^V(\rho)$, if one of the following happens: We have
$(\rho,R^A)\in \C$, or $P$ contains a rule of the form
\[
R^P(\tau)\leftarrow  S_1^P({\sigma_1})\comma
S_2^P({\sigma_2})\comma\dots \comma S_\ell^P({\sigma_\ell}),
\]
where $\tau,\sigma_1,\dots,\sigma_\ell$ are tuples of variables from the set of
variables $X$, and there exists a mapping (evaluation) $\omega\colon X\to V$ such that $\omega(\tau)=\rho$
and for each $i=1,\dots,\ell$ we have $P(I)\proves S_i^V(\omega(\sigma_i))$.

If a Datalog program $P$ ever uses a rule with a goal predicate on its left
side, then the program outputs ``Yes,'' and halts. We will use the symbol $G$
to stand for any of the goal predicates, writing for example $P(I)\proves G$ as
a shorthand for ``$P$ run on $I$ derives a relation that is designated as a
goal predicate.'' Another way to implement goal predicates, used e.g.
in~\cite{feder-vardi}, is to introduce a special nullary relation $G$ that is
the goal. We do not want to deal with nullary relations, but the distinction is
purely a formal one: should the reader want a program with a nullary $G$, all
that is needed is to simply introduce rules of the form $G\leftarrow R(x_1,x_2,\dots)$
where $R$ ranges over the list of goal predicates.

If a goal predicate is not reached, the program $P(I)$ runs until it can not
derive any new statements, at which point it outputs ``No,'' and halts. Thanks
to the monotonous character of Datalog rules (we only add tuples to predicates,
never remove them), any given Datalog program can be evaluated in time
polynomial in the size of its input instance $I$.

Given a Datalog program $P$ and a relational structure $\A$ we say that $P$
\emph{decides $\CSP(\A)$} if $P$ run on a $\CSP(\A)$ instance $I$ reaches a
goal predicate if and only if $I$ is unsatisfiable. We say that $\CSP(\A)$ can
be solved by Datalog if there is a Datalog program $P$
that decides $\CSP(\A)$. (Strictly speaking, we should say that $P$ 
decides $\neg\CSP(\A)$ in this situation, but that is cumbersome.)

For $R_1,\dots,R_k$ relations and $\sigma_1,\dots,\sigma_k$ tuples of
variables, we define the \emph{conjunction} $R_1(\sigma_1) \wedge \dots \wedge
R_k(\sigma_k)$ as a relation (resp. predicate) on $\bigcup_i \Im \sigma_i$.
For example, $R_1(x_3,x_2,x_2)\wedge R_2(x_3,x_4)$ is a relation of arity 3 on
the three variables $x_2,x_3,x_4$.

To slim down our notation, we will for the most part not
distinguish the abstract statement of a Datalog rule (with variables from $X$) and the concrete
realization of the rule (with the evaluation $\omega\colon X\to V$). For example,
if $P$ contained this rule $\alpha$:
\[
  R(x,z)\leftarrow S(x,y)\comma T(y,z)
\]
and it happened that $P(I)\proves S(1,2)$ and $P(I)\proves T(2,2)$, then
instead of saying that we are applying the rule $\alpha$ with the evaluation $\omega(x)=1$, 
$\omega(y)=\omega(z)=2$ to add $(1,2)$ into $R$, we would simply
state that we are using the rule
\[
  R(1,2)\leftarrow S(1,2)\comma T(2,2),
\]
even though that means silently identifying $y$ and $z$ in the original rule
$\alpha$.

The power of Datalog for CSP is exactly the same as that of local consistency
methods. L. Barto and M. Kozik have given several
different natural characterizations of structures $\A$ such that Datalog solves
$\CSP(\A)$~\cite{barto-kozik-bw-2014}. However, this is not the end of the story, for there are natural
fragments of Datalog which have lower expressive power, but also lower
computational complexity.

Predicates that can appear on the left hand side of some rule (and therefore
can have new tuples added into them) are called \emph{intensional database
symbols}
(IDB). Having IDBs on the right hand side of rules enables recursion. Therefore,
limiting the occasions when IDBs appear on the right hand side of rules results
in fragments of Datalog that can be evaluated faster.

An extreme case of such restriction happens when there is never an IDB on the
right hand side of any rule. It is easy to see that such Datalog programs can
solve $\CSP(\A)$ if and only if $\A$ has a finite (also called ``finitary'')
duality, i.e. there exists a finite list $\mathcal Q$ of unsatisfiable $\CSP(\A)$ instances such that
an instance $I$ of $\CSP(\A)$ is unsatisfiable if and only if there exists
$J\in\mathcal Q$ such that one can rename (and possibly also glue together) variables of $J$ to get a subinstance
of $I$. This
property is equivalent to $\CSP(\A)$ being definable in first order logic by
\cite{atserias-duality} (see also the
survey~\cite{bulatov-krokhin-larose-csp-dualities}). Structures of finite
duality are both well understood and rare, so let us look at more permissive
restrictions.

A Datalog program is \emph{linear} if there is at most one IDB on the right
hand side of any rule. When evaluating Linear Datalog programs, we need to only
consider chains of rules that do not branch: It is straightforward to show by
induction that if $P$ is a linear Datalog program and
$I$ is an instance of the corresponding CSP, then
$P(I)\proves R(\rho)$ if and only if $(\rho,R)$ is a constraint of $I$ or there is a sequence of statements
\[
  U_1(\phi_1),U_2(\phi_2),\dots,U_m(\phi_m)=R(\rho)
\]
such that for each $i=2,\dots,m$ the program $P$ has a rule of the form
\[
  U_i(\phi_i)\leftarrow  U_{i-1}({\phi_{i-1}})\comma
T_1^i({\tau_1^i})\comma\dots \comma T_{\ell_i}^i({\tau_{\ell_i}^i}),
\]
where $U_{i-1}$ is the IDB in the rule and $(\tau_j^i,T_j^i)$ are constraints of
$I$ for all
$j=1,\dots,\ell_i$. The first statement, 
$U_1(\sigma_1)$, is a special case as $P$ must derive it without using IDBs, i.e. there is a rule of $P$ of the form
\[
  U_1(\phi_1)\leftarrow  T_1^1(\tau_1^1)\comma\dots \comma
  T_{\ell_1}^1({\tau_{\ell_1}^1})
\]
where all $(\tau_1^1,T_1^1)$, \dots, $(\tau_{\ell_1}^1,T_{\ell_1}^1)$ are
constraints of $I$. 

(Note that this is the first time we are using the ``concrete
realization of the abstract rule'' shorthand.) We will call such a sequence
$U_1(\phi_1),\dots,U_m(\phi_m)$ a \emph{derivation} of $R(\rho)$.

Another way to view the computation of a linear Datalog program is to use the digraph $\mathcal{G}(P,I)$: The set of vertices
of $\mathcal{G}(P,I)$ will consist of all pairs $(\rho,R)$ where $R$ is an
$n$-ary IDB predicate of $P$ and $\rho\in V^n$. The graph $\mathcal{G}(P,I)$
contains the edge from $(\rho,R)$ to $(\sigma, S)$ if $P$ contains a rule of the
form
\[
  R(\rho)\leftarrow  S(\sigma)\comma T_1({\tau_{1}})\comma\dots \comma
  T_k({\tau_{k}}),
\]
where all $(\tau_{i},T_{i})$ are constraints of $I$. 

It is easy to see that $P(I)\proves G$ if and only if there is a tuple $\rho$
and an IDB $R$ such that $P(I)\proves R(\rho)$ in one step, without the use of
intermediate IDBs, and there is a directed path from $(\rho,R)$
to a goal predicate in $\mathcal{G}(P,I)$. It is straightforward to verify that
deciding the existence of such a path is in \compNL{}. (In fact, deciding directed
connectivity is \compNL-complete~\cite[Theorem 4.18,
p.  89]{arora-barak-complexity-textbook} and since there is a linear Datalog
program that decides directed connectivity, it follows that evaluating linear
Datalog programs is \compNL-complete.)

The exact characterization of structures $\A$ such that there is a linear
Datalog program deciding $\CSP(\A)$ is open.  A popular conjecture is that
$\CSP(\A)$ can be solved by linear Datalog if and only if the variety of
$\algA$ admits no tame congruence types except for the lattice and Boolean type
(or equivalently~\cite[Theorem 9.11]{hobby-mckenzie} that $\algA$ generates a
congruence semidistributive variety). 

As Larose and Tesson have shown~\cite[Theorem 4.2]{larose-tesson-hardness-csp},
admitting no types other than lattice and Boolean is necessary for core
relational structures to yield CSPs solvable by linear Datalog. On the other
hand Barto, Kozik and Willard proved that if $\A$ admits an NU polymorphism
then $\CSP(\A)$ can be solved by linear
Datalog~\cite{barto-kozik-willard-nu-linear-datalog}. This is almost, but not
quite, what the necessary condition demands: A finite relational structure $\A$
with finitely many relations has an NU polymorphism if and only if the variety
generated by $\algA$ is congruence distributive~\cite{barto-cd-nu} if and only if the
variety only admits the lattice and/or Boolean types in a particularly nice
way~\cite[Theorem 8.6]{hobby-mckenzie}. 

For our purposes, it will be useful to notice that $\CSP(\A)$ can be solved by
a linear Datalog program if and only if $\A$ has bounded pathwidth duality.

\begin{defi}\label{defPathWidth}
  $\CSP(\A)$ instance $I=(V,\C)$ has \emph{pathwidth} at most $k$ if we can cover 
  $V$ by a family of sets $U_1,\dots,U_m$ such that 
  \begin{itemize}
    \item $|U_i|\leq k+1$ for each $i$,
    \item if $i<j$ and $v\in V$ lies in $U_i$ and $U_j$, then $v$ also lies in
      each of $U_{i+1},\dots,U_{j-1}$, and
    \item for each constraint $C\in \C$ there is an $i$ such that the
  image of the scope of $C$ lies entirely in $U_i$.
  \end{itemize}
\end{defi}

\noindent
The name pathwidth comes from the fact that if we arrange the variables in the
order they appear in $U_1,\dots,U_m$ and look at the instance from far away,
the ``bubbles'' $U_1,\dots, U_m$ form a path. The length of the path is allowed
to be arbitrary, but the ``width'' (size of the bubbles and their overlaps) is
bounded. 

We say that $\A$ has bounded pathwidth duality if there exists a constant $k$
such that for every unsatisfiable instance $I$ of $\CSP(\A)$ there
exists an unsatisfiable instance $J$ of $\CSP(\A)$ of pathwidth at most $k$
such that we can identify some variables of $J$ to obtain a subinstance of $I$.
(This is a translation of the usual definition of duality, which talks about
homomorphisms of relational structures, to CSP instances.)
\begin{prop}[\cite{dalmau-linear-datalog-pw-duality}]
  Assume that $\A$ is a relational structure. Then $\A$ has bounded pathwidth
  duality if and only if there exists a linear Datalog 
  program deciding $\CSP(\A)$.
\end{prop}

Symmetric Datalog is a more restricted version of linear Datalog, where we only
allow \emph{symmetric} linear rules: Any rule with no IDBs on the right
hand side is automatically symmetric, so the interesting case is when a rule
$\alpha$ has the form
\[
  R(\rho)\leftarrow  S(\sigma)\wedge T_1({\tau}_1)\wedge
  T_2({\tau}_2)\wedge\dots,
\]
where $R,S$ are (the only) IDBs. If a symmetric program $P$ contains the rule $\alpha$,
then $P$ must also contain the rule
$\alpha'$ obtained from $\alpha$ by switching $R(\rho)$ and $S(\sigma)$ (we will
call this rule the mirror image of $\alpha$):
\[
  S(\sigma)\leftarrow  R(\rho)\wedge T_1(\tau_1)\wedge
  T_2(\tau_2)\wedge\dots
\]

Observe that if $P$ is a symmetric Datalog program, then $\mathcal{G}(P,I)$ is
always a symmetric graph. Therefore, deciding if $P(I)\proves G$ is equivalent
to an undirected reachability problem. Evaluating symmetric Datalog programs is
thus in $\compL$ thanks to Reingold's celebrated result that undirected
reachability is in \compL~\cite{reingold-reachability}. (In fact, undirected
reachability is \compL-complete under first order
reductions, as is evaluating symmetric Datalog programs: Consider the symmetric
Turing machines introduced in~\cite{sl-paper}. When equipped with logarithmic
amount of memory, these machines define
the complexity class \compSL. When one applies the construction
in the proof of Theorem 3.16 in \cite{descriptive-complexity} to symmetric logspace machines, one gets that
undirected reachability is \compSL-hard modulo first order reductions. Since
$\compL\subset \compSL$~\cite[Theorem 1]{sl-paper}, undirected reachability is $\compL$-hard.)

We will often use Datalog programs whose predicates correspond to relations on
$A$. However, in doing so we will \emph{not} restrict ourselves to just the relations from the relational clone of $\A$. If the predicates
$R^P,S_1^P,\dots,S_\ell^P$ correspond to relations $R^A,S_1^A,\dots,S_\ell^A$
on $A$ in some agreed upon way, then we say that the rule
\[
  R^P(\rho)\leftarrow  S_1^P({\sigma_1})\comma
  S_2^P({\sigma_2})\comma\dots \comma S_\ell^P({\sigma_\ell}),
\]
is \emph{consistent} with $A$ if the
corresponding implication holds for all tuples of $A$, i.e. the sentence 
\[
\forall {f:X\to A},\, R^A(f(\rho))\Leftarrow \left(S_1^A(f(\sigma_1))\wedge
S_2^A(f(\sigma_2))\wedge\dots \wedge S_\ell^A(f(\sigma_\ell))\right),
\]
holds in $A$ (recall that $X$ is the list of all
variables used in the rules of $P$). In other words, a consistent rule records an
implication that is true in $A$.

For $r\in\en$, we construct the \emph{$r$-ary
maximal symmetric Datalog program consistent with $\A$}, denoted by
$\symprogram_\A^{r}$,
as follows: The program has as predicates all relations of arity at most
$r$ on $A$ (these will be IDBs), plus a new symbol for each  
basic relation of $\A$ of arity at most $r$ (these symbols will correspond to the relations used in constraints and
they will never be IDBs; thus we have two symbols for each basic relation of
$\A$, only one of which can be on the left hand side of any rule). 

The set of rules of $\symprogram_\A^{r}$ will contain all rules $\alpha$ that
\begin{enumerate}
  \item are valid linear Datalog rules (i.e. an IDB on the left side, at most one IDB
    on the right),
  \item use only tuples of variables from $X=\{x_1,\dots,x_r\}$ (i.e. at most $r$
    variables at once),
  \item do not have any repetition on the right hand side, i.e. each statement
    $R(\sigma)$ appears in $\alpha$ at most once (however, the predicate $R$ can be used several times with different tuples of variables),
  \item are consistent with $\A$, and
  \item if $\alpha$ contains an IDB on the right hand side, then the mirror
    image $\alpha'$ of $\alpha$ is also consistent with $\A$.
\end{enumerate}
We will designate all empty relations of arity at most $r$ as goal predicates.
We note that our $\symprogram_\A^r$ is a variation of the notion of a canonical 
symmetric Datalog program (used e.g. in~\cite{dalmau-larose-symmetric-datalog}).

It is an easy exercise to show that $\symprogram_\A^r(I)\proves S(\sigma)$ if
and only if $\mathcal{G}(\symprogram_\A^r,I)$ contains a path from $(\rho,A)$ to
$(\sigma,S)$ where $A$ is the unary full relation on $A$ and $\rho$ is arbitrary.
Starting with the full relation will help us simplify proofs by induction later.

The set of rules of $\symprogram_\A^{r}$ is large but finite because there are
only so many ways to choose a sequence of at most $r$-ary predicates on $r$
variables without repetition. Since $\A$ and $r$ are not part of the input of
$\CSP(\A)$, we do not mind that $\symprogram_\A^{r}$ contains numerous
redundant or useless rules.

When we run $\symprogram_\A^{r}$ on a $\CSP(\A)$ instance $I$, it
attempts to narrow down the set of images of $r$-tuples of variables using
consistency:
\begin{obs}\label{obsConsistent}
  Let $\A$ be a relational structure, $r\in\en$, and $I=(V,\C)$ an instance of
  $\CSP(\A)$. Then:
  \begin{enumerate}
    \item\label{itmFirst} if $R^A\subset A^n$, $\rho\in V^n$ are such that
      $\symprogram_\A^{r}(I)\proves R^V(\rho)$, then any solution $f$ of $I$ must 
      satisfy $f(\rho)\in R^A$.
    \item if $\symprogram_\A^{r}(I)\proves G$, then $I$ is not
      satisfiable.
  \end{enumerate}
\end{obs}

\begin{proof}
  To prove the first claim, consider a path in $\mathcal{G}(P,I)$ that witnesses
  $P(I)\proves R(\rho)$:
\[
  (\rho_1,S_1),(\rho_2,S_2),\dots,(\rho_m,S_m)=(\rho,R).
\]
with $S_1^A=A$.

We claim that if $f$ is a solution of $I$, then for each $i=1,\dots,m$ we must
have $f(\rho_i)\in S_i$. We proceed by induction. For $i=1$, this is trivial.

Assume now that $f(\rho_i)\in S_i$ and that $\symprogram_\A^{r}$
contains a rule $\alpha$ of the form
\[
  S_{i+1}(\rho_{i+1})\leftarrow  S_i({\rho_i})\comma
T_1({\tau_1})\comma\dots \comma T_k({\tau_k}),
\]
where $(\tau_j,T_j)\in\C$ for $j=1,\dots,k$. Since $T_j(\tau_j)$ are
constraints of $I$, we have $f(\tau_j)\in T_j$ for each $j$. From the fact that
$\alpha$ is a rule consistent with $\A$, it follows that $f(\rho_{i+1})\in S_{i+1}$.

The second statement of the Lemma is a consequence of the first, since reaching a goal
predicate means that $\symprogram_\A^{r}(I)\proves \emptyset(\rho)$ for some
$\rho$ tuple of variables in $V$. Using
(\ref{itmFirst}), we get that each solution of $I$ must satisfy the impossible
condition $f(\rho)\in \emptyset$ and so there cannot be any solution $f$.
\end{proof}

By Observation~\ref{obsConsistent}, the only way $\symprogram_\A^{r}$
can fail to decide $\CSP(\A)$ is if there is an unsatisfiable instance $I$ of
$\CSP(\A)$ for which $\symprogram_\A^{r}$ does not derive $G$. Our goal in
the rest of the paper is to show that for $r$ large enough and $\A$ nice
enough such a situation will not happen.

Let us close this section by talking about necessary conditions for $\CSP(\A)$
to be solvable by symmetric Datalog. An obvious condition is that, 
since symmetric Datalog is a subset of linear Datalog, $\CSP(\A)$ must be
solvable by linear Datalog.

It turns out that the lower bounds from the tame congruence theory are
compatible with Datalog. If $\A$ is a core, then for $\CSP(\A)$ to be solvable by symmetric
Datalog, $\algA$ must omit all tame congruence theory types
except for the Boolean type~\cite[Theorem 4.2]{larose-tesson-hardness-csp}, from which it
follows~\cite[Theorem 9.14]{hobby-mckenzie} that $\A$ must be variety $n$-permutable for some $n$.
\begin{prop}\label{propOneSide}
  If $\A$ is a core relational structure such that $\CSP(\A)$ is solvable by
  symmetric Datalog, then $\A$ is variety $n$-permutable for some $n$ and $\CSP(\A)$ is
  solvable by linear Datalog.
\end{prop}
Our goal in this paper is to prove that the conditions of
Proposition~\ref{propOneSide} are also sufficient:
\begin{thm}\label{thmSymDatalog}
  Let $\A$ be a relational structure such that there is a linear Datalog
  program that decides $\CSP(\A)$ and $\A$ admits a chain of $n$ Hagemann-Mitschke
  terms as polymorphisms. Then there exists an $r\in \en$ such that
  $\symprogram_\A^{r}$ decides $\CSP(\A)$.
\end{thm}

\section{Stacking symmetric Datalog programs}\label{secStacking}%{{{1 
In this section we describe two tricks that allow us essentially
to run one Datalog program from inside another. The price we pay for this is
that the new program can use fewer variables than the old one.

The first lemma of this section is basically \cite[Lemma
11]{dalmau-larose-symmetric-datalog} rewritten in our
formalism:
\begin{lem}[V. Dalmau, B. Larose]\label{lemConjunction}
  Let $\A$ be a relational structure, $I=(V,\mathcal C)$ an instance of $\CSP(\A)$,
  let $S\subset A^{s}$, $R\subset A^{r}$ be two relations, and let
  $\sigma\in V^{s}$ and $\rho\in V^{r}$. Assume that 
  $\symprogram_{\A}^{s}(I)\proves S(\sigma)$.

  Then for any $k\geq r+s$ we have 
  \[
    \symprogram_{\A}^{k}(I)\proves R(\rho) \Leftrightarrow \symprogram_{\A}^{k}(I)\proves R(\rho)\wedge S(\sigma).
  \]
\end{lem}
\begin{proof}
  Let
  \[
    U_1(\phi_1),U_2(\phi_2),\dots,U_m(\phi_m)=S(\sigma)
  \]
be a path in $\mathcal G(I,\symprogram_\A^s)$ witnessing
  $\symprogram_\A^s(I)\proves S(\sigma)$. Then it is easy to verify that
  \[
   R(\rho), R(\rho)\wedge U_1(\phi_1), R(\rho)\wedge U_2(\phi_2),\dots,
   R(\rho)\wedge U_m(\phi_m)=R(\rho)\wedge S(\sigma)
  \]
  is a path in the graph $\mathcal G(I,\symprogram_\A^k)$. Therefore,
  $\symprogram_\A^k(I)$ derives $R(\rho)$ if an only if it derives $R(\rho)\wedge
  S(\sigma)$.
\end{proof}

Repeated use of Lemma~\ref{lemConjunction} gets us the following:
\begin{cor}\label{corPtimes}
  Let $\A$ be a relational structure, $I$ a $\CSP(\A)$ instance.
  Let $S_1,\dots,S_j$ and $R$ be relations
  on $A$ and $\sigma_1,\dots,\sigma_p,\rho$ be tuples of variables from $I$. 

  If $\symprogram_\A^s(I)\proves S_j(\sigma_j)$ for $j=1,\dots,p$ and  both
  $|\rho|$ and $|\Im \rho\cup\bigcup_{i=1}^p \Im \sigma_i|$ are at most $r$, then we have:
\begin{align*}
  \symprogram_\A^{r+s}(I)\proves R(\rho)\Leftrightarrow &
  \symprogram_\A^{r+s}(I)\proves R(\rho)\wedge S_{1}(\sigma_1)\\
  \Leftrightarrow&
  \symprogram_\A^{r+s}(I)\proves R(\rho)\wedge S_{1}(\sigma_1)\wedge
 S_{2}(\sigma_2)\\
  &\vdots\\
 \Leftrightarrow&
 \symprogram_\A^{r+s}(I)\proves R(\rho)\wedge S_{1}({\sigma_{1}})\wedge\dots\wedge
    S_{p}({\sigma_{p}})
  \end{align*}
\end{cor}

\begin{defi}
Given an instance $I=(V,\C)$ of $\CSP(\A)$, we say that $\symprogram^r_{\A}$ derives
the instance $J=(W,\D)$ from $I$, writing $\symprogram^r_\A(I)\proves J$, if $W\subset V$ 
and for each $(\sigma,R)\in \D$ we have  $\symprogram^r_\A (I)\proves R(\sigma)$.
\end{defi}

Obviously, if $\symprogram^r_\A$ derives an unsatisfiable instance from $I$,
then $I$ itself is unsatisfiable. Moreover, a maximal
symmetric Datalog program run on $I$ can simulate the run of a smaller maximal
symmetric Datalog program on $J$:

\begin{lem}\label{lemCompose1}
  Let $\A=(A,R_1,\dots,R_n)$ and $\B=(A,S_1,\dots,S_m)$ be two relational structures 
  and let $I=(V,\mathcal{C})$ be an instance of
  $\CSP(\A)$. Assume that $r,s$ are positive integers and $J=(W,\mathcal{D})$
  is an instance of $\CSP(\B)$ such that $\symprogram_\A^s(I)\proves J$ and $\symprogram_\B^r(J)\proves G$.
  Then $\symprogram_{\A}^{r+s}(I)\proves G$.
\end{lem}

\begin{proof}
  The derivation of $\symprogram_\A^{r+s}(I)\proves G$ will follow the
  derivation $\symprogram_{\B}^r(J)\proves G$, generating the constraints of $J$
  on the fly using $\symprogram_\A^{s}(I)$. Note that since $\A$ and $\B$ share 
  the same base set, the predicates of $\symprogram_\B^r$ are also predicates of
  $\symprogram_\A^{r+s}$.

  Let $U_1(\phi_1),U_2(\phi_2),\dots,U_q(\phi_q)$ be a derivation of $G$ by
  $\symprogram_{\B}^{r}(J)$ such that $U_1=A$. 

   We proceed by induction on $i$ from $1$ to $q$ and show that
   $\symprogram_{\A}^{r+s}(I)\proves U_i(\rho_i)$ for all $i$. Since all goal predicates of
   $\symprogram_\A^r$ are also goal predicates of $\symprogram_\A^{r+s}$, this
   will show that $\symprogram_{\A}^{r+s}(I)\proves G$.
   The base case is easy: Since $U_1$ is full, $\symprogram_\A^{r+s}$ has the
   rule ``$U_1(\phi_1)\leftarrow \quad$'', giving us $\symprogram_\A^{r+s}(I)\proves U_1(\phi_1)$.

   Assume that $\symprogram_{\A}^{r+s}(I)\proves U_i(\phi_i)$.  
   Since $\symprogram_{\B}^{r}(J)$ derives $U_{i+1}(\phi_{i+1})$ from
   $U_{i}(\phi_i)$, there have to be numbers $j_1,\dots,j_p$ and tuples
   $\sigma_1,\dots,\sigma_p$ such that each $(\sigma_k,S_{j_k})$ is a constraint of
   $J$, and
   \[
     U_{i+1}(\phi_{i+1}) \leftarrow  U_i(\phi_i)\comma S_{j_1}(\sigma_{1})\comma
   S_{j_2}(\sigma_{2})\comma\dots\comma S_{j_p}(\sigma_{p})
 \]
 is a rule of $\symprogram_{\B}^{r}$. From this, it is easy to verify that the following
 rule, which we will call $\alpha$, is a rule of $\symprogram_\A^{r+s}$:
  \begin{align*}
    (U_{i+1}({\phi_{i+1}})\wedge S_{j_1}({\sigma_{1}})\wedge\dots\wedge
    S_{j_p}({\sigma_{p}})) &\leftarrow  \\
    (U_i({\phi_i})&\wedge
    S_{j_1}({\sigma_{1}})\wedge\dots\wedge
    S_{j_p}({\sigma_{p}})),
  \end{align*}

    Since $\symprogram_\A^{r+s}(I)\proves U_i(\phi_i)$, 
    Corollary~\ref{corPtimes} yields $\symprogram_\A^{r+s}(I)\proves
    U_i(\phi_i)\wedge \bigwedge_{k=1}^p S_{j_k}({\sigma_{k}})$. We then use
    the rule $\alpha$ to obtain 
    $\symprogram_\A^{r+s}(I)\proves
    U_{i+1}(\phi_{i+1})\wedge \bigwedge_{k=1}^p S_{j_k}({\sigma_{k}})$ and
    finally use the other implication from Corollary~\ref{corPtimes} to get
$\symprogram_\A^{r+s}(I)\proves
    U_{i+1}(\phi_{i+1})$, concluding the proof.
\end{proof}

At one point, we will need to look at powers of $\A$. For this, we introduce the following notation: If
\[
  \sigma=((s_{1,1},\dots,s_{k,1}),\dots,(s_{\ell, 1},\dots,s_{\ell,k}))\in
(A^k)^\ell
\]
is an $\ell$-tuple of elements of $A^k$ then by $\overline \sigma$
we will mean the $k\ell$-tuple we get by ``unpacking'' $\sigma$ into $A^{k\ell}$:
\[
\overline\sigma=(s_{1,1},\dots,s_{k,1},\dots,s_{\ell, 1},\dots,s_{\ell,k}).
\]
If $U\subset (A^k)^\ell$ is a relation on $A^k$, we will denote by
$\overline U$ the relation $\overline U=\{\overline \sigma\colon \sigma\in
U\}\subset A^{k\ell}$.

The following lemma generalizes Lemma~\ref{lemCompose1} to powers of
$A$. The proof is similar to that of Lemma~\ref{lemCompose1} and we omit it for brevity.

\begin{lem}\label{lemCompose2}
  Let $k\in\en$ and assume we have relational structures $\A$ and $\B$ on the
  sets $A$ and $A^k$ respectively. Assume moreover that $I=(V,\mathcal{C})$ is an instance of
  $\CSP(\A)$, $S_1,\dots,S_m$ are basic relations of $\B$,
  $\sigma_1,\dots,\sigma_m$ are tuples of elements
  of $V^k$, and $r,s$ are positive integers such that:
  \begin{enumerate}
    \item $\symprogram_\A^{r}(I)\proves \overline{S}_i(\overline{\sigma_i})$ for each
      $i=1,\dots,m$,
    \item $\symprogram_{\B}^{s}(J)\proves G$, where 
      $J$ is the instance $J=(V^k,\{(\sigma_i,S_i)\mid i=1,\dots,m\})$ of
      $\CSP(\B)$.
  \end{enumerate}
  Then $\symprogram_{\A}^{r+ks}(I)\proves G$.
\end{lem}

\section{Variety \texorpdfstring{$n$}{n}-permutability on path instances}\label{secNPermutable}%{{{1

We begin our construction by showing how variety $n$-permutability limits the kind of
CSP instances a symmetric Datalog program can encounter.

\begin{defi}
  An instance $I=(V,\mathcal C)$ of CSP is a \emph{path instance} of length $\ell$ if:
  \begin{enumerate}
    \item $V$ is a linearly ordered set (we use $V=[\ell]$ ordered by size whenever
      practicable, such as in the rest of this definition),
    \item for each $i\in V$, $I$ contains exactly one unary constraint with 
      scope $i$; we will denote its constraint relation by $B_i\subset A$,
    \item for each $i=1,2,\dots,\ell-1$, $I$ contains exactly one binary constraint
      with scope $(i,i+1)$; we denote its constraint relation $B_{i,i+1}$.
    \item $I$ contains no other constraints than the ones named above.
  \end{enumerate} 
\end{defi}
\noindent Note that $B_{i,i+1}$ can contain tuples from outside of
$B_i\times B_{i+1}$. We allow that to happen to simplify our later arguments.

\begin{figure}
  \begin{center}
    \begin{tikzpicture}[bend angle=45,draw,auto,node distance=1cm]
      \node[vertex] (a1){};
      \node[vertex] (b1)[below of=a1, node distance=2cm]{};
      \foreach \i/\j in {1/2,2/3,3/4} {
	\node[vertex] (a\j)[right of=a\i, node distance=4cm]{};
	\node[vertex] (b\j)[right of=b\i, node distance=4cm]{};
      }
      \foreach \i in {1,2,3,4} {
	\node (aa\i) [above of=a\i] {};
	\node (bb\i) [below of=b\i, label=right:$i_\i$] {};
        \draw[dashed] (aa\i)--(bb\i);
      }
      \foreach \i/\j/\k in {0/1/2,1/2/3,2/3/4} {
	\draw (b\j)--(b\k) node [midway,label=below:$s_\i$]{};
	\draw (a\j)--(b\k) node [midway]{$s_\j$};
	\draw (a\j)--(a\k) node [midway,label=above:$s_\k$]{};
      } 
      \draw[decoration={zigzag,segment length=2mm,amplitude=1mm},decorate] (b1) to[out=0, in=100](a4);
      \node[above left of=a4, node distance=2cm]{$t$};
  \end{tikzpicture}
  \end{center}
  \caption{A sketch of a 4-braid. The solution $t$ from Observation~\ref{obsZZ}
    pictured as a zigzag.}
  \label{figBraid}
\end{figure}
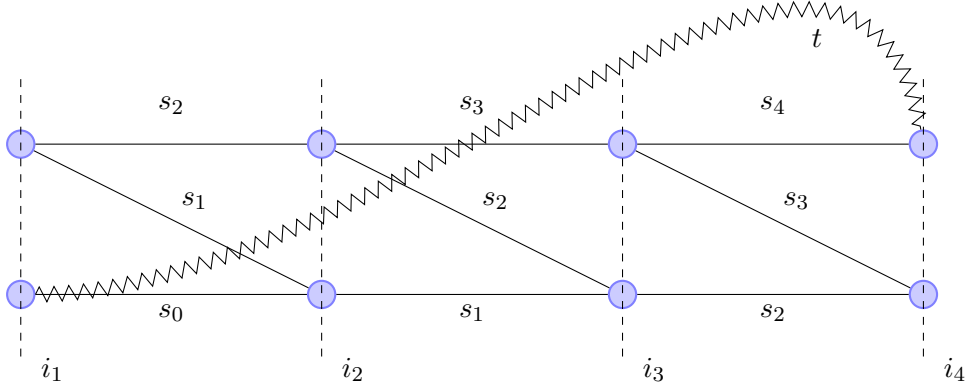

If $I$ is a path instance of length $\ell$ and $a\leq b$ are integers, we define
the instance $I$ restricted to $[a,b]$ as the
subinstance of $I$ induced by all variables of $I$ from the $a$-th
to the $b$-th (inclusive). We will denote $I$ restricted to $[a,b]$ by $I_{[a,b]}$.
\begin{defi}
  Let $I$ be a path CSP instance on $[\ell]$ and $n\geq 2$ be an integer. An \emph{$n$-braid} (see
  Figure~\ref{figBraid}) in $I$ is a collection
  of $n+1$ solutions $s_0,s_1,s_2,\dots,s_n$ of $I$ together with indices
  $1\leq i_1<\dots<i_n\leq \ell$ such that for all $k=1,2,\dots,n-1$ we have
  \begin{enumerate}
    \item $s_k(i_k)=s_{k+1}(i_k)$, and
    \item $s_{k-1}(i_{k+1})=s_{k}(i_{k+1})$.
  \end{enumerate}
  When we want to explicitly describe a braid, we will often give the $(2n+1)$-tuple
  \[
    (s_0,s_1,\dots,s_n;i_1,\dots,i_n).
    \]
\end{defi}

We care about braids because it is easy to apply Hagemann-Mitschke terms to
them to get new solutions of $I$. This observation 
is not new; one can find it formulated in a different language
in~\cite[Theorem 8.4]{freese-valeriote-maltsev-conditions}:
\begin{obs}[R. Freese, M. Valeriote]\label{obsZZ}
  Let $n\in\en$ and let $\algA$ be a variety $n$-permutable algebra,
  $I$ be a path instance of $\CSP(\algA)$, and let
  $(s_0,\dots,s_n;$ $i_1,\dots,i_n)$ be an $n$-braid in $I$. Then there exists a solution
  $t$ of $I$ such that $t(i_1)=s_{0}(i_1)$ and $t(i_n)=s_{n}(i_n)$.
\end{obs}
\begin{proof}
  Since $\algA$ is variety $n$-permutable, we have a chain of Hagemann-Mitschke terms
  $p_0,p_1,\dots,p_n$ compatible with constraints of $I$. All we need to do is
  apply these terms  on $s_0,s_1,\dots,s_n$.

  Denote by $r_k$ the mapping $r_k(i)=p_k(s_{k-1}(i),s_{k}(i),s_{k+1}(i))$
  where $k$ goes from 1 to $n-1$; we let $r_0=s_0$ and $r_n=s_n$. Since $p_k$
  is a polymorphism, each $r_k$ is a solution of $I$. Moreover, one can verify
  using the Hagemann-Mitschke equations together with the equalities from the
  definition of an $n$-braid that for each $k=1,\dots,n$ we have
  $r_{k-1}(i_k)=r_{k}(i_k)$.

  Since $I$ is a path instance, we can glue the solutions $r_0,\dots,r_n$
  together: The mapping $t$ defined as $t(i)=r_k(i)$ whenever $i_k<i\leq
  i_{k+1}$ (where we put $i_{-1}=0$ and $i_{n+1}=\ell$ for convenience)
  is a solution of $I$. To finish the proof, it remains to observe that $t(i_1)=s_0(i_1)$ and
  $t(i_n)=s_n(i_n)$.
\end{proof}

Let $I$ be a path instance of CSP. We will say that a binary constraint
$B_{i,i+1}$ of $I$ is \emph{subdirect} if $B_i,B_{i+1}\neq\emptyset$, $B_i\subset
\pi_1(B_{i,i+1}\cap B_i\times B_{i+1})$, and
$B_{i+1}\subset \pi_2(B_{i,i+1}\cap B_i\times B_{i+1})$. (We have modified the standard definition of
subdirectness a bit to account for the fact that $B_{i,i+1}$ can
contain tuples outside of $B_i\times B_{i+1}$.) An instance is subdirect if all
its constraints are subdirect. Observe that every a subdirect path instance is 
satisfiable.
\begin{obs}
If $I$ is a subdirect
path instance and $e\in
(B_i\times B_{i+1})\cap B_{i,i+1}$, then by walking from $e$ backwards and
  forwards along the edges defined by the binary constraints of $I$ we get a
  solution $s$ of $I$ that contains
the edge $e$, that is $(s(i),s(i+1))=e$.
\end{obs}

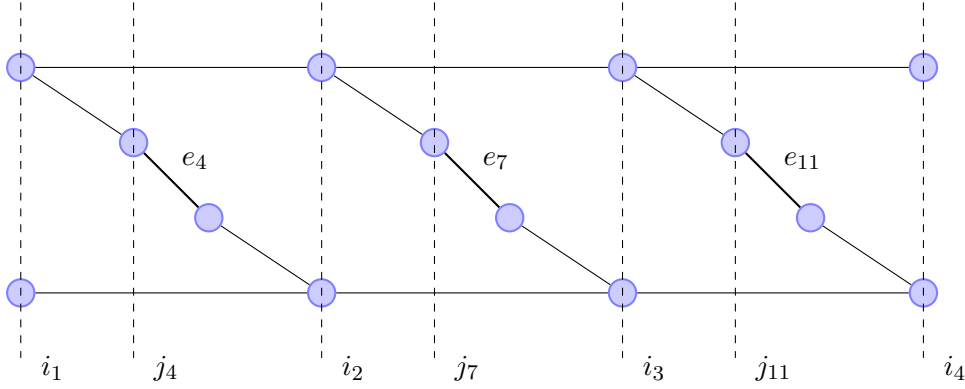
\begin{figure}
  \begin{center}
    \begin{tikzpicture}[bend angle=45,draw,auto,node distance=1cm]
      \node[vertex] (a1){};
      \node[vertex] (b1)[below of=a1, node distance=3cm]{};
      \foreach \i/\j in {1/2,2/3,3/4} {
	\node[vertex] (a\j)[right of=a\i, node distance=4cm]{};
	\node[vertex] (b\j)[right of=b\i,node distance=4cm]{};
	\node[vertex] at ($(a\i)+(1.5cm,-1cm)$) (c\i) {};
	\node[vertex] at ($(c\i)+(1cm,-1cm)$) (d\i) {};
      }
      \foreach \i in {1,2,3,4} {
	\node (aa\i) [above of=a\i] {};
	\node (bb\i) [below of=b\i, label=right:$i_\i$] {};
        \draw[dashed] (aa\i)--(bb\i);
      }

      \foreach \i/\j in {1/4,2/7,3/11} {
	\node at ($(c\i)+(0cm,2cm)$) (ctop\i) {};
	\node[label=right:$j_{\j}$] at ($(c\i)+(0cm,-3cm)$) (cbottom\i) {};
        \draw[dashed] (ctop\i)--(cbottom\i);
	\draw[thick] (c\i)--(d\i) node [midway]{$e_{\j}$};
      }
      \foreach \i/\j in {1/2,2/3,3/4} {
	\draw (b\i)--(b\j);
	\draw (a\i)--(c\i);
	\draw (d\i)--(b\j);
	\draw (a\i)--(a\j);
      }
  \end{tikzpicture}
  \end{center}
  \caption{The conclusion of Lemma~\ref{lemBraidExists} for $n=3$. Important
  edges $e_i$ drawn in
  bold.}
  \label{figBraidExists}
\end{figure}

The following lemma tells us that if a path instance $I$ is subdirect and we mark enough edges
in $I$, we can find an $n$-braid that goes through many edges of our choosing.
It is a Ramsey-like result and we prove it using the Ramsey theorem (see e.g.~\cite[Theorem 3.3]{a-course-in-combinatorics}).
\begin{lem}\label{lemBraidExists}
  For every $n$ and $N$ there exists an $m$ with the following property: Let 
  $I$ be a subdirect path CSP
  instance of length $\ell>m$ such that $|B_i|\leq N$ for each $i\in[\ell]$. Then for any
  choice of indices $1\leq j_1<j_2<\dots<j_m<\ell$ and edges $e_k\in
  B_{j_k,j_k+1}\cap (B_{j_k}\times B_{j_k+1})$
  for $k=1,\dots,m$, there exists
  an $n$-braid $(s_0,\dots,s_n; i_1,\dots,i_n)$ in $I$ 
 such that for every
  $k=1,2,\dots,n-1$ there is a
  $q$ so that $i_k\leq j_q < i_{k+1}$  and $(s_k(j_q),s_k(j_q+1))=e_q$ (that is,
  between every pair of ``crossings'' is an edge $e_q$; see
  Figure~\ref{figBraidExists}).
\end{lem}

\begin{proof} 
  Without loss of generality we can assume that $B_i\subset [N]$ for each $i$. For each $k=1,\dots,m$, we
  choose and fix a solution $\sigma_k$ of $I$ that contains the edge $e_k$
  (which we get from subdirectness of $I$; see above). 

  Consider now the complete graph $G$ with vertex set $[m]$ whose edges are
  colored as follows: For every $u<v$ we color the edge $\{u,v\}\in
  \binom{m}{2}$ by the pair of
  numbers $(\sigma_u(j_v),\sigma_v(j_u))\in [N]^2$. By the Ramsey
  theorem, if $m$ is large enough then there exists a monochromatic induced
  subgraph of $G$ on $2n+1$ vertices. To make our notation simpler, we will
  assume that these vertices are $1,2,\dots,2n+1$. 

  Thanks to edges of $G$ being monochromatic on $[2n+1]$, we have that $\sigma_u$ and $\sigma_{u'}$ agree on $j_v$ as long as
  $u,u',v\in [2n+1]$ and either $u,u'<v$, or $u,u'>v$. Using this, we can easily
  verify that $(\sigma_1,\sigma_3,\dots,\sigma_{2n+1};j_2,j_4,\dots,j_{2n})$ is
  an $n$-braid. For each $k=1,2,\dots,n-1$ we get:
  \begin{align*}
    \sigma_{2k+1}(j_{2k})&=\sigma_{2k+3}(j_{2k})\\
    \sigma_{2k+1}(j_{2k+2})&=\sigma_{2k-1}(j_{2k+2})
  \end{align*}
  To finish the proof, observe that for every $k\in [n-1]$ we have $j_{2k}<j_{2k+1}<j_{2k+2}$ and the 
  solution $\sigma_{2k+1}$ was chosen so that it passes through $e_{2k+1}$, so
  we can let $q=2k+1$ and satisfy the conclusion of the lemma.
\end{proof}

Given a path instance $I$, we will define the sets $C_i\subseteq B_i$ by $C_1=B_1$
and 
\[
  C_{i+1}=\{b\in B_{i+1}\colon \exists c\in C_i, (c,b)\in B_{i,i+1}\}.
  \]

The sets $C_i$ correspond to the endpoints of solutions of $I_{[1,i]}$, so $I$ 
is satisfiable if and only if $C_\ell\neq\emptyset$. We will
call an edge $(d,c)\in B_{i,i+1}$ such that $d\in B_i\setminus C_i$ and $c\in C_{i+1}$ a
\emph{backward edge}.  

Our goal in Section~\ref{secSymDLPaths} will be to show how to use symmetric Datalog to
identify unsatisfiable path $\CSP(\A)$ instances for $\A$ fixed and variety $n$-permutable.
We will see that in the absence of backward edges a simple symmetric Datalog program can identify
all unsatisfiable path CSP instances. This is why we want to know what happens
when there are many backward edges.  It turns out that an 
variety $n$-permutable instance that has too many backward edges is never subdirect. In
Section~\ref{secSymDLPaths}, this will
enable us to reduce the size of the instance.

\begin{lem}\label{lemBWedges}
  For every $n$ and $N$ there exists an $m$ such that if $I$ is a path instance
  of length $\ell>m$ and $1<a<b<\ell$ are such that 
  \begin{enumerate}
    \item each set $B_i$ has cardinality at  most $N$, and
    \item all sets $B_i$ and all relations $B_{i,i+1}$ are invariant under a chain of $n$
      Hagemann-Mitschke terms, and
    \item there are at least $m$ distinct indices $j$ in $[a,b)$ such that $B_{j,j+1}$
	contains a backward edge,
   \end{enumerate}
   then the instance $I_{[a,b]}$ is not subdirect.
\end{lem}

\begin{proof}
  We pick $m$ large enough to be able to use Lemma~\ref{lemBraidExists} for
  sets $B_i$ of maximum size $N$ and $(n+1)$-braids. Taking this $m$, we look at 
  what would happen were $I_{[a,b]}$ subdirect. 

  Let $a\leq j_1<\dots<j_m<b$ be a list of indices where backward edges occur in
  $[a,b)$. For each $k=1,\dots,m$, we choose a backward edge $e_{j_k} \in
  B_{j_k,j_k+1}$ and apply Lemma~\ref{lemBraidExists} to $I_{[a,b]}$. We obtain an $(n+1)$-braid 
  in $I_{[a,b]}$ that uses $n+1$ of our backward edges; denote the solutions
  and indices that make up this
  braid by $s_0,\dots,s_{n+1}$ and $i_1,i_2,\dots,i_{n+1}$, respectively. Moreover, since $s_1$ passes through a backward edge $e_j$ for some
  $j\in [i_1,i_2)$, we get $s_1(i_2)\in C_{i_2}$.  Since the only condition on
  $s_0$ is $s_0(i_2)=s_1(i_2)$, we can modify $s_0$ to ensure $s_0(i_1)\in
  C_{i_1}$ without breaking the braid. The situation is sketched in
  Figure~\ref{figBridgingGap}.

  \begin{figure}
  \begin{center}
    \begin{tikzpicture}[bend angle=45,draw,auto,node distance=1cm]
      \node[vertex] (a1){};
      \node[vertex] (b1)[below of=a1, node distance=3cm]{};
      \foreach \i/\j/\k in {1/2/3,2/3/4,3/4/3} {
	\node[vertex] (a\j)[right of=a\i, node distance=\k cm]{};
	\node[vertex] (b\j)[right of=b\i,node distance=\k cm]{};
	\node[vertex] at ($(a\i)+(1cm,-1cm)$) (c\i) {};
	\node[vertex] at ($(c\i)+(1cm,-1cm)$) (d\i) {};
      }

      \node [left of=b1] {$s_0(i_1)$};
      \node [left of=a1] {$s_1(i_1)$};
      \node [above of=a3] {$s_n(i_n)$};
      \draw [dashed] (b1) to[out=0,in=200](a3);

      \foreach \i/\j in {1/1,2/2,3/n,4/n+1} {
	\node[below of=b\i]{$B_{i_{\j}}$};
      }
      \foreach \i in {0,1,2,3,4,5,7,8,9,10} {
	\draw (\i,-2.4) ellipse (0.4cm and 1cm);
	\draw
	($(\i,0)+(-0.4,0.4)$)--($(\i,0)+(0.4,0.4)$)--($(\i,-3)+(0.4,-0.4)$)--($(\i,-3)+(-0.4,-0.4)$)--($(\i,0)+(-0.4,0.4)$);
      }
     
      \foreach \i in {1,2,3} {
	\node at ($(c\i)+(0cm,2cm)$) (ctop\i) {};
	\node at ($(c\i)+(0cm,-3cm)$) (cbottom\i) {};
	\draw[very thick] (c\i)--(d\i) node [midway] {};
      }
      \node [right of=c3, node distance=8mm] {$e_j$};
      \foreach \i/\j in {1/2,3/4} {
	\draw (b\i)--(b\j);
	\draw (a\i)--(c\i);
	\draw (d\i)--(b\j);
	\draw (a\i)--(a\j);
      }
      \node[left of=a3] (aa3) {$\dots$};
      \node[left of=b3] (bb3) {$\dots$};
      \draw (a2)--(aa3);
      \draw (a2)--(c2);
      \draw (d2)--(bb3);
      \draw (b2)--(bb3);
      \draw (bb3)--(b3);
      \draw (aa3)--(a3);
      
  \end{tikzpicture}
    \end{center}
    \caption{A schematic view of the instance $I_{[a,b]}$ (the ellipses are the
    sets $C_i$, backward edges $e_j$ are thick).}
  \label{figBridgingGap}
\end{figure}
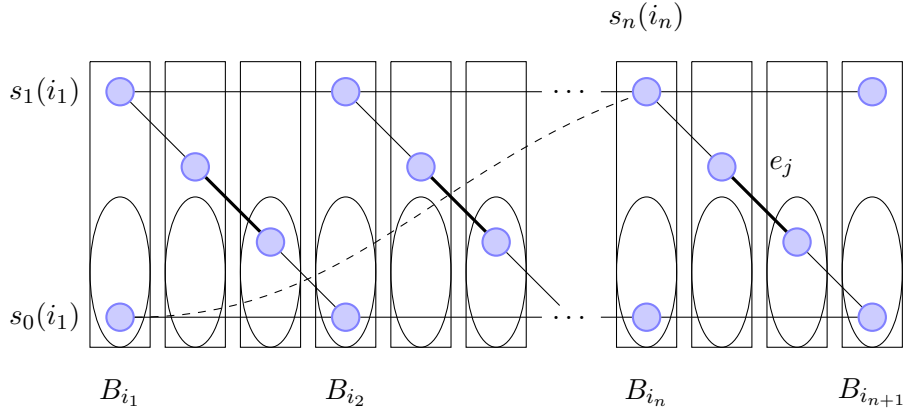
  Observation~\ref{obsZZ} then gives us that 
  $I_{[a,b]}$ has a solution $t$ such that $t(i_1)=s_{0}(i_1)\in C_{i_1}$ 
  and $t(i_n)=s_{n}(i_n)=s_{n+1}(i_n)$ (shown by a dashed line in
  Figure~\ref{figBridgingGap}).

  Now it remains to see that since $t(i_n)=s_{n+1}(i_n)$, there is a path
  from $t(i_n)$ to some backward edge $e_j$, $j\geq i_{n+1}$. Therefore,
  $t(i_n)\in B_{i_n}\setminus C_{i_n}$ and solution $t$ witnesses that
  there is a path from $t(i_1)\in C_{i_1}$ to $t(i_n)\not\in C_{i_n}$, a
  contradiction with the way we have defined the sets $C_i$.
\end{proof}

\section{Undirected reachability on path instances}\label{secLambda}%{{{1
Given a path CSP instance $I$, we
define the digraph $\Conn(I)$ of $I$ as the directed graph with vertex set equal to the
disjoint union of all unary constraints $B_1,\dots,B_n$ and edge set equal to the disjoint union of
all binary constraints of $I$ (restricted to the sets $B_i$). The orientation of
$\Conn(I)$ establishes levels on the graph ($B_1$ is on the first level, $B_{2}$ on the
  second level and so on).

Given a path CSP instance $I$ and numbers $i\leq j$, the relation
$\lambda_{I,i,j}$ consists of all pairs $a\in B_i,b\in B_j$ that lie in the
same component of weak connectivity of  $\Conn({I_{[i,j]}})$ (i.e.
there
is an oriented, but not necessarily directed, path from $a$ to $b$ in
$\Conn(I_{[i,j]})$).

\begin{lem}\label{lemPPdef}
  If $I$ is a path CSP instance of $\CSP(\A)$ and $i\leq j$, then $\lambda_{I,i,j}$ lies in the
  relational clone of $\A$.
\end{lem}
\begin{proof}
  It is easy to see that for $a\in B_i$ and $b\in B_j$ we have
  $(a,b)\in\lambda_{I,i,j}$ if and only if there is a digraph homomorphism
  $h\colon P\to \Conn({I_{[i,j]}})$ where $P$ is an oriented path which starts at level 0,
  ends at level $j-i$, has no vertex of level less than 0 or more than
  $j-i$, and $h$ maps the starting point of $P$ to $a$ and ending point of $P$
  to $b$. 

  Let now the path $P$ witness $(a,b)\in\lambda_{I,i,j}$ and the path $Q$
  witness $(c,d)\in \lambda_{I,i,j}$.   By~\cite[Lemma 2.36]{hell-nesetril},
  $P\times Q$ then contains an oriented path $R$ that goes from level 0 to
  level $j-i$. By considering projections of $P\times Q$, we obtain that $R$
  homomorphically maps to both $P$ and $Q$ and from this it
  is easy to verify that $R$ witnesses both $(a,b),(c,d)\in\lambda_{I,i,j}$.
  Since there are only finitely many pairs in $\lambda_{I,i,j}$, we can repeat
  this procedure to find a path $S$ that witnesses the whole $\lambda_{I,i,j}$.
  It is then straightforward to translate homomorphisms from $S$ to $\Conn({I_{[i,j]}})$ into a 
  primitive positive definition of $\lambda_{I,i,j}$ in $\A$.  
\end{proof}

\begin{lem}\label{lemTwoStones}
  For every relational structure $\A$, every path instance $I$ of $\CSP(\A)$,
  and every $i\leq j$, we have $\symprogram_{\A}^{3}(I)\proves \lambda_{I,i,j}(i,j)$.
\end{lem}
\begin{proof}
  Let us fix $i$ and $j$. For $k\in\{i,i+1,\dots,j\}$, consider
  the relation 
  \begin{align*}
    \rho_{k}=\{(a,b)\in B_i\times B_k\colon& \text{$a$, $b$ lie in the same
    component}\\
    &\text{of weak connectivity of $\Conn(I_{[i,j]})$}\}.
  \end{align*}
  We show by induction on $k$ that $\symprogram_{\A}^{3}(I)\proves \rho_k(i,k)$ for every
  $k=i,\dots,j$. This will be enough, since $\rho_{j}=\lambda_{I,i,j}$.

  The base case $k=i$ is easy: Since $\rho_i \supseteq \{(b,b)\colon b\in
  B_i\}$, the program $\symprogram_{\A}^{3}$ contains the rule $\rho_i(x,x)\leftarrow 
  B_i(x)$, so we get
  $\symprogram_{\A}^3(I)\proves \rho_i(i,i)$.

  The induction step: Assume we have $\symprogram_\A^3\proves\rho_{k}(i,k)$.
  Given the definition of $\rho_k$ and $\rho_{k+1}$, it is straightforward to
  verify that the pair of rules
  \begin{align*}
    \rho_{k+1}(x,z) &\leftarrow \rho_{k}(x,y)\wedge B_{k,k+1}(y,z) \\
    \rho_{k}(x,y) &\leftarrow \rho_{k+1}(x,z)\wedge B_{k,k+1}(y,z) 
  \end{align*}
  is consistent with $\A$ and therefore present in $\symprogram_\A^3$.
  Applying the first of those rules (with $x=i$, $y=k$, and $z=k+1$) then gives
  us $\symprogram_\A^3(I)\proves \rho_{k+1}(i,k+1)$, completing the proof. 
\end{proof}

Let $I$ be a path instance of $\CSP(\A)$ of length $\ell$. In the following, we 
will again be using the sets $C_i$ from Section~\ref{secNPermutable}.

Let $1<i_1<i_2<\dots<i_k< \ell$ be the complete list of all indices $i$ with a
backward edge in $B_{i,i+1}$ (i.e. all $i$ such that
that $B_{i,i+1}\cap\left((B_i\setminus C_i)\times C_{i+1}\right)\neq\emptyset$).
For convenience, we let $i_0=0$ and $i_{k+1}=\ell$. 

\begin{figure}
  \begin{center}
    \begin{tikzpicture}[bend angle=45,draw,auto,node distance=1cm]
	\draw (1,-2.4) ellipse (0.4cm and 1cm);
	\draw
	($(1,-1.8)+(-0.4,0.4)$)--($(1,-1.8)+(0.4,0.4)$)--($(1,-3)+(0.4,-0.4)$)--($(1,-3)+(-0.4,-0.4)$)--($(1,-1.8)+(-0.4,0.4)$);
	\node [vertex] at (1,-2)  (c1) {};
	\node [vertex] at (1,-3)  (cc1) {};
	\node at (1,-1) (B1) {$B_{1}$};

      \foreach \i/\j in {3/4,6/7,10/11} {
	\draw (\i,-2.4) ellipse (0.4cm and 1cm);
	\draw (\j,-2.4) ellipse (0.4cm and 1cm);
	\node [vertex] at (\i,-1) (b\i) {};
	\node [vertex] at (\i,0) (bb\i) {};
	\node [vertex] at (\j,0) (bb\j) {};
	\node [vertex] at (\i,-2)  (c\i) {};
	\node [vertex] at (\i,-3)  (cc\i) {};
	\node [vertex] at (\j,-2)  (c\j) {};
	\node [vertex] at (\j,-3)  (cc\j) {};
	\draw ($(\i,0)+(-0.4,0.4)$)--($(\i,0)+(0.4,0.4)$)--($(\i,-3)+(0.4,-0.4)$)--($(\i,-3)+(-0.4,-0.4)$)--($(\i,0)+(-0.4,0.4)$);
	\draw ($(\j,0)+(-0.4,0.4)$)--($(\j,0)+(0.4,0.4)$)--($(\j,-3)+(0.4,-0.4)$)--($(\j,-3)+(-0.4,-0.4)$)--($(\j,0)+(-0.4,0.4)$);
	\node at (\i,0.8) (B\i) {$B_{\i}$};
	\node at (\j,0.8) (B\j) {$B_{\j}$};
	\draw (b\i)--(c\j);
	\draw (c\i)--(cc\j);
	\draw (bb\i)--(bb\j);
	\node at ($(\i,-3.8)!0.5!(\j,-3.8)$) (B\i\j) {$B_{\i,\j}$};
      }
      \foreach \i/\j/\amp in {1/3/5} {
	\coordinate (tmp1) at ($(c\i)!0.5!(c\j)$);
	\coordinate (tmp2) at ($(cc\i)!0.5!(cc\j)$);
	\node[below of=tmp2] (lambda1) {$\lambda_{I,\i,\j}$};
	\draw [decoration={zigzag, segment length=2mm, amplitude= \amp mm},
	decorate] (tmp1)-- (tmp2);
      }

      \foreach \i/\j/\amp in {4/6/5,7/10/10} {
      \coordinate (tmp1) at ($(c\i)!0.5!(c\j)$);
      \coordinate (tmp2) at ($(cc\i)!0.5!(cc\j)$);
      \coordinate (tmp3) at ($(tmp1)+(0,2)$);
      \coordinate (tmp4) at ($(tmp2)+(0,2)$);
      \node[below of=tmp2] (lambda1) {$\lambda_{I,\i,\j}$};
      \draw [decoration={zigzag, segment length=2mm, amplitude= \amp mm}, decorate] (tmp1)-- (tmp2);
      \draw [decoration={zigzag, segment length=2mm, amplitude= \amp mm}, decorate] (tmp3)-- (tmp4);
    }
     \end{tikzpicture}
  \end{center}
  \caption{The instance $I_\lambda$ with $i_1=3,i_2=6,i_3=10$ (ellipses mark
  the sets $C_i=D_i$).}
  \label{figIlambda}
\end{figure}
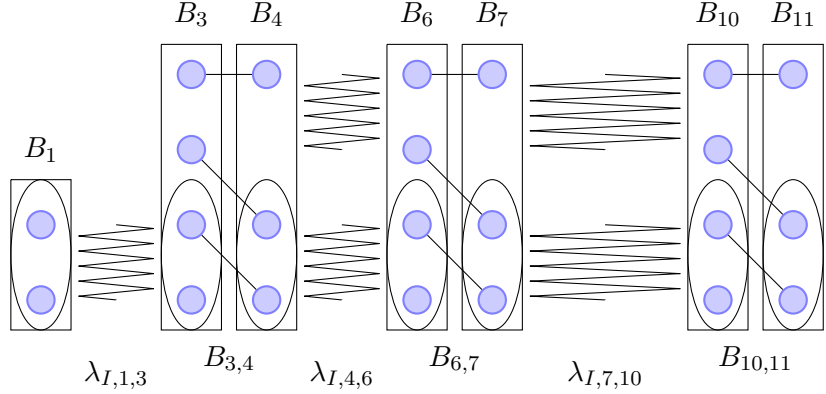

Now consider the new path instance $I_{\lambda}$ (see Figure~\ref{figIlambda})
with variable set 
\[
  U=\{1,i_1,i_1+1,i_2,\dots,i_k,i_k+1,\ell\}.
\]  
We get
$I_\lambda$ from $I_{\upharpoonright U}$ by filling out the gaps by relations
$\lambda_{I,i_j+1,i_{j+1}}$: For all $j$ such that
$i_j+1 < i_{j+1}$ (i.e. $I_{\upharpoonright U}$ has no binary constraint between $i_j+1$ and
$i_{j+1}$), we add the binary constraint
$((i_j+1,i_{j+1}),\lambda_{I,i_j+1,i_{j+1}})$ to $I_\lambda$.  See Figure~\ref{figIlambda}.

By Lemma~\ref{lemPPdef}, the constraints of $I_\lambda$ belong to the relational
clone of $\A$. Let for each $v\in U$ the set $D_{v}\subset
B_v$ consist of all values of $s(v)$ where $s$ is a solution of $(I_\lambda)_{[1,v]}$.
It is easy to show by induction on $v$ that $D_v=C_v$ for all $v\in U$. In particular, we have that $I_\lambda$ is satisfiable if 
and only if $I$ is satisfiable. Moreover, $I_{\lambda}$ has a backward edge in
roughly every other binary constraint. Finally, $\symprogram_\A^3$
derives $I_{\lambda}$ from $I$ by Lemma~\ref{lemTwoStones}.

We can summarize the findings of this section as follows:

\begin{lem}\label{lemLinked}
  Let $\A$ be a relational structure and let 
  $I$ be an unsatisfiable path instance of $\CSP(\A)$. Then $\symprogram_\A^3$
  derives from $I$ the unsatisfiable path instance $I_\lambda$ with the
  following property: For all $m\geq 1$, any
  interval of variables of $I_\lambda$ of length at least $2m+2$ contains at
  least $m$ indices with backward edges and the
  constraints of $I_\lambda$ are invariant under all polymorphisms of $\A$.
\end{lem}

\section{Symmetric Datalog solves all variety \texorpdfstring{$n$}{n}-permutable path instances}%{{{1
\label{secSymDLPaths} In this section, we put together the results
from the previous two sections to show that for every variety $n$-permutable $\A$ there
is an $M$ such that $\symprogram_\A^M(I)\proves G$ for every unsatisfiable path
instance $I$ of $\CSP(\A)$: 

\begin{thm}\label{thmSolveLinear}
  For each $N$ and $n$ there exists $f(n,N)\in \en$ so that whenever 
  $\A$ is a variety $n$-permutable relational structure and $I$ an 
  unsatisfiable path instance of $\CSP(\A)$ such that
  $|B_i|\leq N$ for all $i$, then $\symprogram_\A^{f(n,N)}(I)\proves G$.
\end{thm}
\begin{proof}
  We prove the theorem first in the case when $\A$ contains symbols for all
  binary and unary relations compatible with $\algA$, and then show how the general
  case follows.

  We fix $n$ and proceed by induction on $N$. For $N=1$, a path instance is unsatisfiable
  if and only if at least one of $B_{i,i+1}$ does not intersect $B_i\times
  B_{i+1}$, which $\symprogram_\A^2$
  easily detects, so $f(n,1)=2$ works.

  Assume that the theorem is true for all structures and all instances with
  sets $B_i$ smaller than $N$.  Let $m$ be the number from Lemma~\ref{lemBWedges}
  for our $n$ and $N$. We let $f(n,N)=f(n,N-1)+2m+6$ and claim that
  $\symprogram_\A^{f(n,N)}(I)\proves G$ for any $I\in\CSP(\A)$ whose unary
  constraints $B_i$ have at most $N$ elements. For brevity, let us denote $2m+2$
  by $L$, so we have $f(n,N)=f(n,N-1)+L+4$.

  Our starting point is the instance $I_\lambda$ from Section~\ref{secLambda}.
  By the first part of Lemma~\ref{lemLinked}, $\symprogram_\A^3(I)\proves
  I_{\lambda}$ and $I_\lambda$ is an unsatisfiable path CSP instance of
  $\CSP(\A)$. Consider now what $\symprogram_\A^{L+1}$ does on $I_\lambda$.
  First of all, if the length of $I_\lambda$ is at most $L$, then
  $\symprogram_\A^{L+1}$ can easily check feasibility of $I_\lambda$ by looking
  at the whole instance at once. So if $I_\lambda$ is short, we get
  $\symprogram_\A^{L+1}(I_\lambda)\proves G$ and we are done (by
  Lemma~\ref{lemCompose1}, we have $\symprogram_\A^{L+4}(I)\proves G$). This is
  why in the rest of the proof we will assume that $I_\lambda$ is longer than $L$. We show that
  $\symprogram_\A^{L+1}(I_\lambda)$ derives another unsatisfiable instance $K$
  that falls within the scope of the induction hypothesis.

  It turns out that $I_\lambda$
  contains many backward edges: By Lemma~\ref{lemLinked}, each interval of $I_\lambda$
  of length $2m+2$ contains at least $m$ backward edges. We can thus
  use Lemma~\ref{lemBWedges} to show that any interval of $I_\lambda$ of length
  $L$ contains at least one binary constraint that is not subdirect. These
  constraints will enable us to shrink the unary constraints on $I_\lambda$.

 Let $\ell$ be the length of $I_\lambda$. For $1\leq a\leq i\leq b\leq
  \ell$ we will introduce the following two relations:
  \begin{align*}
    S_{I_\lambda,[a,b],i}&=\{s(i)\colon \text{$s$ is a solution of
    ${\left(I_\lambda\right)}_{[a,b]}$}\},\\
    S_{I_\lambda,[a,b]}&=\{(s(a),s(b))\colon \text{$s$ is a solution of
    ${\left(I_\lambda\right)}_{[a,b]}$}\}. 
  \end{align*}
  It is easy to see that these relations lie in the relational clone of $\A$.
  From the definitions above, it easily follows that 
  $\symprogram_\A^{L+1}(I_\lambda)\proves S_{I_\lambda,[a,b]}(a,b)$ and 
  $\symprogram_\A^{L+1}(I_\lambda)\proves S_{I_\lambda,[a,b],i}(i)$
  whenever $b-a\leq L$ (this can be done in one step as the program is big
  enough to simply look at the whole of $(I_\lambda)_{[a,b]}$
  at once).

We are now ready to show that $\symprogram_\A^{L+1}(I_\lambda)\proves K$, where
$K$ is an unsatisfiable path instance of $\CSP(\A)$ whose unary constraints all
have at most $N-1$ elements. 
 
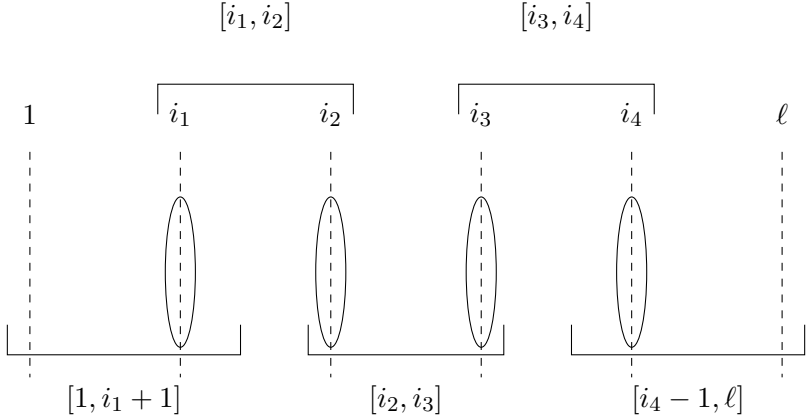
\begin{figure}
  \begin{center}
    \begin{tikzpicture}[bend angle=45,draw,auto,node distance=1cm]
      \foreach \i/\j in {2/1,4/2,6/3,8/4} {
	\draw (\i,-2.4) ellipse (0.2cm and 1cm);
	\node at (\i,-0.3) (B\i) {${i_\j}$};
	\draw[dashed] (\i,-0.8)--(\i,-3.8);
      }
	\node at (0,-0.3) (B0) {1};
	\draw[dashed] (-0,-0.8)--(-0,-3.8);
	\node at (10,-0.3) (Bl) {$\ell$};
	\draw[dashed] (10,-0.8)--(10,-3.8);

      \foreach \i/\j/\k/\l in {0/2.5/1/i_1+1,4/6/i_2/i_3,7.5/10/i_4-1/\ell} {
	\draw
	($(\i,-3.1)+(-0.3,0)$)--($(\i,-3.1)+(-0.3,-0.4)$)--($(\j,-3.1)+(0.3,-0.4)$)--($(\j,-3.1)+(0.3,0)$);
	\node at ($(\i,-4.2)!0.5!(\j,-4)$) (i\i\j) {$[\k,\l]$};
      }
      \foreach \i/\j/\k/\l in {2/4/1/2,6/8/3/4} {
	\draw
	($(\i,-0.3)+(-0.3,0)$)--($(\i,-0.3)+(-0.3,0.4)$)--($(\j,-0.3)+(0.3,0.4)$)--($(\j,-0.3)+(0.3,0)$);
	\node at ($(\i,1)!0.5!(\j,1)$) (i\i\j) {$[i_\k,i_\l]$};
      }
  \end{tikzpicture}
    \end{center}
    \caption{Constructing the instance $K$ by looking at solutions of intervals
      of $I_\lambda$. Unary constraints of $K$ shown as ellipses.}
  \label{figGettingK}
\end{figure}
  We construct $K$ as follows: Denote by $B_i'$ the unary constraint on the
  $i$-th variable of $I_\lambda$. Since subdirectness fails somewhere in 
  $[1,L]$, there is an index $i_1\in
  [1,L]$ such that $S_{I_\lambda,[1,i_1+1],i_1}$ is strictly smaller than
  $B_{i_1}'$. Looking at $[i_1+1,i_1+L]$, we find an index
  $i_2$ where subdirectness fails again, so
  $S_{I_\lambda,[i_2-1,i_2+1],i_2}\subsetneq B_{i_2}'$. After that, we find 
  $i_3\in [i_2+1,i_2+L]$ such that $S_{I_\lambda,[i_3-1,i_3+1],i_3}\subsetneq
  B_{i_3}'$ and so on. 
  Continuing in this manner, we get an 
  increasing sequence of indices $1\leq i_1<i_2<\dots<i_k\leq \ell$ such that for
  each $j=2,\dots,k-1$ we have $|S_{I_\lambda,[i_j-1,i_j+1],i_j}|\leq |B_{i_j}'|-1$ and
  $i_{j+1}-i_j\leq L$. We end this process when $i_k+L>\ell$. To
  properly analyze what goes on at the end of the chain, we need to consider two cases: 
  When $i_k<\ell$ and when $i_k=\ell$.
  It is straightforward to verify that in both cases we have $\ell-i_k+1\leq L$ and 
  $S_{I_\lambda,[i_k-1,\ell],i_k}\leq |B_{i_k}'|-1$.

  We take these indices $i_j$ and observe that we have the following derivations
  (see Figure~\ref{figGettingK} for reference; note that $L$ is at least 2):
  \begin{align*}
    \symprogram_\A^{L+1}(I_\lambda)&\proves S_{I_\lambda,[i_j-1,i_j+1],i_j}(i_j)
    \quad\text{for all $j=2,\dots,k-1$,}\\
  \symprogram_\A^{L+1}(I_\lambda)&\proves S_{I_\lambda,[1,i_1+1],i_1} (i_1),\\
    \symprogram_\A^{L+1}(I_\lambda)&\proves
    S_{I_\lambda,[i_{k}-1,\ell],i_k}(i_k),\\
\symprogram_\A^{L+1}(I_\lambda)&\proves S_{I_\lambda,[i_1,i_2]} (i_1,i_2),\\
\symprogram_\A^{L+1}(I_\lambda)&\proves S_{I_\lambda,[i_2,i_3]} (i_2,i_3),\\
        &\vdots\\
\symprogram_\A^{L+1}(I_\lambda)&\proves
    S_{I_\lambda,[i_{k-1},i_k]} (i_{k-1},i_k).
  \end{align*}
  We take these relations and use them to build up our instance $K$ of
  $\CSP(\A)$: The instance $K$ has variables $i_1,i_2,\dots,i_k$. The
  constraints of $K$ are as follows: $K$ has unary constraints 
  $(i_1,S_{I_\lambda,[1,i_1+1],i_1})$ (for the first variable), 
  $(i_j,S_{I_\lambda,[i_j-1,i_j+1],i_j})$ for $j=2,\dots,k-1$, and 
  $(i_k,S_{I_\lambda,[i_k-1,\ell],i_k})$ for the last variable. The binary
  constraints of $K$ are 
$((i_j,i_{j+1}),S_{I_\lambda,[i_j,i_{j+1}]})$ where $j=1,\dots,k-1$.

  Since the relations $S_{I_\lambda,\dots}$ incorporate all constraints of
  $I_\lambda$, it is straightforward to see that any solution of $K$ would give us a solution of
  $I_\lambda$, so $K$ is unsatisfiable. Moreover, all unary constraints of $K$ have
  at most $N-1$ members and all constraint relations of $K$
  belong to the relational clone of $\A$. By the induction hypothesis, we then have
  $\symprogram_\A^{f(n, N-1)}(K)\proves G$. It now remains to use
  Lemma~\ref{lemCompose1} twice: We get first
  $\symprogram_\A^{f(n,N-1)+L+1}(I_\lambda)\proves G$, followed by 
  $\symprogram_\A^{f(n,N-1)+L+4}(I)\proves G$. Since we chose $f(n,N)$
  to be $f(n,N-1)+L+4$, we have the desired result $\symprogram_\A^{f(n,N)}(I)\proves G$.

  It remains to talk about the case when $\A$ does not contain symbols for
  all unary and binary compatible relations. Denote by $\B$ the relational
  structure we get from $\A$ by adding those missing relational symbols. 
  Let $I$ again be an instance of $\CSP(\A)$ with each $B_i$ of size at most $N$. 
  By the above argument, we get $\symprogram_\B^{f(n,N)}(I)\proves G$, so there is a derivation of $G$ in
  $\symprogram_\B^{f(n,N)}$ from the relations of $I$. Observe now that the
  instance $I$ only contains relations from $\A$ and that if we take
  $\symprogram_\B^{f(n,N)}$ and delete rules that contain non-IDB predicates
  (the name used in the literature for non-IDB predicates is extensional database symbols) that are not basic
  relations of $\A$, we get $\symprogram_\A^{f(n,N)}$. Therefore, the derivation
  of $\symprogram_\B^{f(n,N)}(I)\proves G$ also
  witnesses that $\symprogram_\A^{f(n,N)}(I)\proves G$ and we
  are done. 
\end{proof}

By taking $M=f(n,|A|)$, we obtain the following corollary:
\begin{cor}\label{cor:path-n-perm}
 For each variety $n$-permutable relational structure $\A$ there exists $M\in \en$ so that whenever 
$I$ is an 
  unsatisfiable path instance of $\CSP(\A)$, then $\symprogram_\A^{M}(I)\proves
  G$.
\end{cor}

\section{From linear to symmetric Datalog}\label{secFinale}%{{{1
It remains to explain how to move from solving path CSP instances to
solving general CSP instances. This is where we will need linear Datalog, or
equivalently bounded pathwidth duality.

Given a relational structure $\A$, we use the idea from~\cite[Proposition
13]{Barto-kozik-approximation} and define the $k$-th \emph{bubble power} of
$\A$ as the structure $\A^{(k)}$ with the universe $A^k$ and the following basic
relations:
\begin{enumerate}
  \item All unary relations $S\subset A^k$ that can be
    defined by taking a conjunction of basic relations of $\A$ (we are also
    allowed to identify variables and introduce dummy variables, but not to
    do existential quantification), and
   \item all binary relations of the form
     \[
       E_{\mathcal I}=\left\{((a_1,\dots,a_k),(b_1,\dots,b_k))\in
     \left(A^k\right)^2 \colon \forall (i,j)\in \mathcal I,\, a_i=b_j\right\}
   \]
     where
     $\mathcal I\subset [k]^2$.
\end{enumerate}
In this section, we show that if $\A$ has pathwidth duality at most $k-1$,
then all we need to worry about are path CSP instances of $\CSP(\A^{(k)})$. Our
method is straightforward, but we need to get a bit technical to take care of
all details.

\begin{lem}\label{lemAlmostThere}
Let $\A$ be a (finite) relational structure, $k\in \en$. Assume that
$\A$ has pathwidth duality $k-1$ and let $s\in\en$ be such that $\symprogram_{\A^{(k)}}^{s}(I)\proves G$ 
for each unsatisfiable path instance $I$ of $\CSP(\A^{(k)})$. Then 
$\symprogram_{\A}^{k(s+2)}$ decides $\CSP(\A)$.
\end{lem}
\begin{proof}
  We need to show that $\symprogram_\A^{k(s+2)}(I)\proves G$ for
  every unsatisfiable instance $I$. Since $\A$ has pathwidth duality $k-1$, it
  is enough to show that $\symprogram_\A^{k(s+2)}(J)\proves G$ whenever $J=(V,\C)$ is an
  unsatisfiable $\CSP(\A)$ instance of pathwidth at most $k-1$.

  Let $X_1,\dots,X_\ell$ be the partition
  of $V$ witnessing that $J$ has pathwidth at most $k-1$. If $X_i\subset X_{i+1}$
  resp. $X_{i+1}\subset X_i$ for some $i$, then we can delete the smaller of
  the two sets and still have a partition that
  satisfies Definition~\ref{defPathWidth}. Therefore, we can assume that all
  neighboring sets are incomparable. From this, it follows that all sets $X_i$
  are pairwise different, because $X_i=X_j$ for $i<j$ implies $X_i\subset X_{i+1}$.

  We fix a linear order $\prec$ on $V$. 
  For each $i$, we will represent $X_i$ by the $k$-tuple $\chi_i\in X_i^k$ that
  we get by listing the elements of $X_i$ from $\prec$-minimal to
  $\prec$-maximal, repeating the $\prec$-maximal element if $X_i$ has less than
  $k$ elements. Since the sets $X_i$ are pairwise different, we get pairwise
  different tuples. Recall that $J_{\upharpoonright X_i}$ denotes the subinstance of $J$
  induced by $X_i$.

  We now construct an unsatisfiable path instance $K$ of $\CSP(A^{(k)})$. The variable
  set of $K$ is $\{\chi_1,\dots,\chi_\ell\}$. The constraints are as follows:
  \begin{enumerate}
    \item   For each $i$, the $i$-th unary constraint relation $B_i$ lists
      all solutions of $J_{\upharpoonright X_i}$. More
      formally, we let
      \[
	B_i=\{ \rho\circ\chi_i \colon \rho \in A^{X_i},\,\text{is a solution of
	$J_{\upharpoonright X_i}$}\}\subset A^k.
      \]
      It is straightforward to verify that $B_i$ is a basic relation of $\A^{(k)}$.

    \item For each $i=1,2,\dots,\ell-1$, we encode the intersection of $X_i$
      and $X_{i+1}$ by adding the constraint $B_{i,i+1}=E_{\mathcal I}$ where
      $\mathcal I=\{(a,b)\colon \chi_i(a)=\chi_{i+1}(b)\}$.
  \end{enumerate}
  If $r$ is a solution of $K$, we can construct a solution $t$ of $J$ as
  follows: For each $v\in V$, find an $i\in[\ell]$ and $j\in[k]$ such that $\chi_i(j)=v$ 
  and let $t(v)$ be the $j$-th coordinate of $r(\chi_i)$. It is an easy
  exercise to verify that the $t$ we obtain would be a solution of $J$.
  Since $J$ is unsatisfiable, so is $K$.
  
  Since $K$ is a path instance, we get $\symprogram_{\A^{(k)}}^{s}(K)\proves G$.
  Now extend the set of variables of $K$ to the whole
  $V^k$ without adding any new constraints. While this new instance $K'$ is no
  longer a path instance, it is still true that
  $\symprogram_{\A^{(k)}}^{s}(K')\proves G$ (the derivation of $G$ can just
  ignore the new variables).

  We can now use Lemma~\ref{lemCompose2}: The structure $\B$ in the Lemma
  will be $\A^{(k)}$ and the relations $S_1,\dots,S_m$ will be 
  $B_1,B_2,\dots,B_\ell$ and $B_{1,2},B_{2,3},\dots,B_{\ell-1,\ell}$. It is
  straightforward to show that $\symprogram^{2k}(J)$ derives the instance $K$:
  Each of the statements
  $\symprogram^{2k}(J)\proves \overline{B_i}(\overline{\chi_i})$ and
  $\symprogram^{2k}(J)\proves
  \overline{B_{i,i+1}}(\overline{\chi_i},\overline{\chi_i})$ (where $i$ ranges
  over $[\ell]$ and $[\ell-1]$, respectively) has a derivation of length one. 
  Lemma~\ref{lemCompose2} then gives us that 
  $\symprogram_{\A}^{ks+2k}(J)\proves G$, concluding the proof.
\end{proof}

We are now ready to prove our main result:

\begin{thm*}[Theorem \ref{thmSymDatalog} restated]
  Let $\A$ be a relational structure such that there is a linear Datalog
  program that decides $\CSP(\A)$ and $\A$ admits a chain of $n$ Hagemann-Mitschke
  terms as polymorphisms. Then there exists a number $M$ so that
  $\symprogram_\A^{M}$ decides $\CSP(\A)$.
\end{thm*}
\begin{proof}
  Since there is a linear Datalog program that decides $\CSP(\A)$, there is a
  $k\in\en$ so that $\A$ has pathwidth duality at most $k$. 

  It is straightforward to verify that the basic relations of the bubble power $\A^{(k)}$ are compatible with
  the Hagemann-Mitschke terms of $\A$ applied componentwise (recall that the
  universe of $\A^{(k)}$ is the $k$-th power of $A$), so $\A^{(k)}$ is
  variety $n$-permutable. Corollary~\ref{cor:path-n-perm} then gives us that there is an integer $M'$ such
  that the program   $\symprogram_{\A^{(k)}}^{M'}$ derives the goal predicate on any unsatisfiable path
  instance of $\CSP(\A^{(k)})$. Therefore, Lemma~\ref{lemAlmostThere} gives us
  that $\symprogram_\A^{(k+2)M'}$ decides $\CSP(\A)$.
\end{proof}
\section{Conclusions}%{{{1
In Theorem~\ref{thmSymDatalog}, we gave a characterization of the class of CSPs
solvable by symmetric Datalog programs. Unfortunately, our result depends on
understanding the power of linear Datalog; the
characterization of CSPs solvable by linear Datalog is an
open problem at the moment.

However, once somebody obtains a characterization of linear Datalog, our result
immediately gives a characterization of symmetric Datalog. To see how that could
come about, let us reexamine some conjectures about the CSPs solvable by
fragments of Datalog~\cite{larose-tesson-hardness-csp} that would give us a
characterization of symmetric Datalog:

\begin{conj}[B. Larose, P. Tesson]\label{conjSDjoin}
  Let $\A$ be a finite relational structure such that the algebra of polymorphisms of
  $\A$ generates a variety that only admits the lattice and/or Boolean tame
  congruence theory types (equivalently, the variety is congruence
  semidistributive). Then there is a linear
  Datalog program that decides $\CSP(\A)$.
\end{conj}

An alternative way to settle the complexity of CSPs solvable by symmetric
Datalog would be to replace ``linear Datalog'' in Theorem~\ref{thmSymDatalog}
by just ``Datalog''. In particular, if the following were true, we would get 
a characterization of symmetric Datalog, too:

\begin{conj}[B. Larose, P. Tesson]\label{conjType3}
  Let $\A$ be a relational structure such that the algebra of polymorphisms
  $\algA$ of $\A$ is idempotent and generates a variety that only admits the Boolean  
  tame congruence theory type. Then $\CSP(\A)$ is solvable by linear Datalog.
\end{conj}

If Conjecture~\ref{conjSDjoin} or~\ref{conjType3} is true,
then the following are
equivalent for any core relational structure $\A$:
\begin{enumerate}
  \item $\A$ is variety $n$-permutable for some $n$ and
    $\CSP(\A)$ is solvable by Datalog.\label{itmNpermBW}
  \item The idempotent reduct of $\algA$ generates a variety that admits only the tame congruence theory
    type~3.\label{itmPure3}
  \item There exists a symmetric Datalog program that decides
    $\CSP(\A)$.\label{itmExistsSymDL}
\end{enumerate}
Here the implication $(\ref{itmNpermBW})\Rightarrow(\ref{itmExistsSymDL})$ (or
$(\ref{itmPure3})\Rightarrow(\ref{itmExistsSymDL})$) is
the unknown one. Implication
$(\ref{itmExistsSymDL})\Rightarrow(\ref{itmPure3})$ follows from~\cite[Theorem
4.2]{larose-tesson-hardness-csp}, while~\cite[Theorem 9.15]{hobby-mckenzie} together
with the characterization of problems solvable by
Datalog~\cite{barto-kozik-bw-2014} gives us $(\ref{itmNpermBW})\Leftrightarrow
(\ref{itmPure3})$.

We end with another citation of~\cite{larose-tesson-hardness-csp} whose
consequences we find tantalizing: Assume that $\compL\neq \compNL$ and $\compL\neq\compModL{p}$ for any
$p$ prime. Then we can add a fourth statement to the above list:
\begin{enumerate}
  \item[(d)]$\CSP(\A)$ is in $\compL$ modulo first order reductions.
\end{enumerate}
From one side,
symmetric Datalog programs can be evaluated in logarithmic space. For the other
implication, we cite Theorem~\ref{thmTypeLB} to see that unless $\algA$ only
admits the Boolean type, there is a first order reduction to $\CSP(\A)$ from
a problem that is $\compNL$-hard or $\compModL{p}$-hard for some
$p$.
\section*{Acknowledgments}
I would like to thank Ralph McKenzie for support and mentoring. I'm grateful to
Pascal Tesson for helping me understand first order reductions better and to
Benoit Larose for pointing out an error in the definition of subdirect
constraints in the preprint of this paper. Finally,
I thank the referees for their perceptive and valuable comments.

This work was begun at Vanderbilt University, Nashville, TN, USA,
and was finished at IST Austria and Charles University. Funded by European Research Council 
under the European Unions Seventh Framework Programme (FP7/2007-2013)/ERC 
grant agreement no. 616160 and PRIMUS/SCI/12 project of the Charles University.

\bibliographystyle{alpha}%{{{1
\bibliography{citations}
\end{document}